\documentclass[sigconf]{acmart}
\newcommand{\revised}[1]
{\textcolor{black}{#1}}
\usepackage{hyperref}
\usepackage{multirow}
\usepackage{hhline}
\usepackage{graphicx}
\usepackage{array}

\usepackage{longtable}
\usepackage{booktabs}

\hypersetup{
    colorlinks=true,
    linkcolor=blue,
    filecolor=magenta,      
    urlcolor=cyan,
}
\AtBeginDocument{%
  }

\copyrightyear{2026}
\acmYear{2026}
\setcopyright{rightsretained}
\acmDOI{10.1145/3675094.3678379}

\usepackage{amsmath}
\usepackage{booktabs}
\usepackage{multirow}
\usepackage{amsmath,amsfonts}
\usepackage{algorithmic}
\usepackage{algorithm}
\usepackage{array}
\usepackage[caption=false,font=normalsize,labelfont=sf,textfont=sf]{subfig}
\usepackage{textcomp}
\usepackage{stfloats}
\usepackage{url}
\usepackage{verbatim}
\usepackage{graphicx}
\usepackage{hhline}
\usepackage{makecell}
\usepackage{enumitem}
\usepackage{tabularx}
\newcolumntype{Y}{>{\centering\arraybackslash}X} 
\usepackage{enumitem}

\acmConference[CHI '26]{CHI}{April 13–17,
  2026}{Barcelona, Spain}
\acmBooktitle{CHI (CHI '26), April 13–17, 2026, Barcelona, Spain}
\acmISBN{0000}

\begin{document}


\title[From Daily Song to Daily Self]{From Daily Song to Daily Self: Supporting Reflective Songwriting of Deaf and Hard-of-Hearing Individuals through Generative Music AI}


\author{Youjin Choi}
\orcid{0000-0003-2788-7871}
\authornote{First author.}
\affiliation{%
  \institution{Department of AI Convergence, Gwangju Institute of Science and Technology}
  \city{Gwangju}
  \state{Buk-gu}
  \country{Korea, Republic of,}
}
\email{chldbwls304@gm.gist.ac.kr}

\author{JinYoung Yoo}
\orcid{0009-0002-0751-4349}
\affiliation{%
  \institution{Department of AI Convergence, Gwangju Institute of Science and Technology}
  \city{Gwangju}
  \state{Buk-gu}
  \country{Korea, Republic of,}
}
\email{9inceroo@gm.gist.ac.kr}

\author{JaeYoung Moon}
\orcid{0000-0002-1852-2769}
\affiliation{%
  \institution{Department of AI Convergence, Gwangju Institute of Science and Technology}
  \city{Gwangju}
  \state{Buk-gu}
  \country{Korea, Republic of,}
}
\email{super_moon@gm.gist.ac.kr}

\author{Yoonjae Kim}
\orcid{0009-0005-6174-2483}
\affiliation{%
  \institution{Department of AI Convergence, Gwangju Institute of Science and Technology}
  \city{Gwangju}
  \state{Buk-gu}
  \country{Korea, Republic of,}
}
\email{rladbswo12@gm.gist.ac.kr}

\author{Eun Young Lee}
\orcid{}
\affiliation{%
  \institution{Department of Music Therapy, Ewha Womans University}
  \city{Seoul}
  \country{Korea, Republic of,}
}
\email{eleemt@ewha.ac.kr}

\author{Jennifer G Kim}
\authornote{Co-senior authors contributed equally to this research.}
\orcid{0000-0003-3253-3963}
\affiliation{%
  \institution{School of Interactive Computing, Georgia Institute of Technology}
  \city{Atlanta}
  \state{Georgia}
  \country{United States}
}
\email{jennifer.kim@cc.gatech.edu}

\author{Jin-Hyuk Hong}
\authornotemark[2]
\authornote{Corresponding author.}
\orcid{0000-0002-8838-5667}
\affiliation{%
  \institution{Department of AI Convergence, Gwangju Institute of Science and Technology}
  \city{Gwangju}
  \state{Buk-gu}
  \country{Korea, Republic of,}
}
\email{jh7.hong@gist.ac.kr}


\renewcommand{\shortauthors}{Choi et al.}

\begin{abstract}
The rapid advancement of generative AI (GenAI) is expanding access to songwriting, offering a new medium of self-expression for Deaf and Hard-of-Hearing (DHH) individuals. \revised{However, emerging technologies that support DHH individuals in expressing themselves through music have largely been evaluated in single-session settings and often fall short in helping users unfamiliar with songwriting convey personal narratives or sustain engagement over time.} This paper explores songwriting as an extended, \revised{music-based journaling practice that supports sustained emotional reflection over multiple sessions.} We introduce \textit{SoulNote}, a GenAI system enabling DHH to engage in iterative songwriting. Grounded in user-centered design, including a design workshop, a preliminary study, and \revised{a multi-session diary study}, our findings show that ongoing songwriting with \textit{SoulNote} facilitated emotional growth across three dimensions: self-insight, emotion regulation, and \revised{everyday attitudes toward emotions and self-care}. Overall, this work demonstrates how GenAI can support marginalized communities by transforming creative expression into a daily practice of self-discovery and reflection.


\end{abstract}




\maketitle

\section{Introduction}
Communication constraints in an auditory-centered society can contribute to psychological challenges for Deaf and Hard-of-Hearing (DHH) individuals, including social isolation, reduced self-esteem, and negative self-perception \cite{Goldblat2018, Davidsson2018, Punch2004, Olsson2021}. Music-based activities have gained attention as a promising means of supporting DHH individuals’ psychological well-being by enabling self-expression \cite{HyeonBeom2025, Ward2016, Small1998}. \revised{Songwriting, in particular, allows individuals to embed personal experiences into lyrics and musical elements, supporting self-expression and identity reconstruction} \cite{Youjin2025, ChungHa2025, soderberg2016}. Music generative AI (GenAI) offers new opportunities for creating songs that reflect individual narratives \cite{Youjin2025CHI2, Louie2020}, \revised{and text-based prompting has lowered technical barriers to music creation for DHH users} \cite{Youjin2025CHI}. \revised{Recent work on music GenAI, however, has raised substantial concerns about authorship and creative ownership, algorithmic bias, environmental cost, and potential emotional risks for users \cite{de2025towards, fivser2025emotional}. According to the guidelines for trustworthy AI music systems proposed by Jacopo \textit{et al.} \cite{de2025towards}, music GenAI can raise issues related to the copyright status of training data and the possibility that generated pieces resemble copyrighted works, as well as the risk of producing lyrics containing offensive or harmful language and sonic characteristics that may induce anxiety or distress. Despite these potential issues, we leverage music GenAI in this work because it provides an accessible pathway for DHH individuals to explore songwriting and represent their lived experiences through music.}

However, conveying personal narratives through music remains challenging for DHH individuals who are relatively less familiar with music \cite{Ziqing2025, Rezwana2023}. \revised{Existing studies using music GenAI have primarily examined short-term accessibility and emotional benefits \cite{ChungHa2025, Youjin2025CHI}, yet DHH users still often struggle to articulate emotions in lyrics and to translate them into musical forms that reflect their intended meanings. The lack of interactive and sustained prompting partly explains why prior work has focused on single-session experiences \cite{Youjin2025, ChungHa2025, Youjin2025CHI}. Research has not yet investigated how repeated or long-term songwriting may support ongoing emotional reflection. Although single-session activities can promote temporary emotional awareness, they are insufficient for capturing the dynamics of emotional processing} \cite{James1997, Felicity2013}. Evidence from journaling and psychotherapy highlights the importance of repeated self-expression for mental health \cite{James1997, Pennebaker2011, Allison2011}, and therapeutic songwriting research suggests that multi-session engagement can support emotional healing and self-reflection \cite{Eickholt2022, Gee2019}. Extending songwriting into a longitudinal, journaling-like practice may offer deeper and more sustained emotional benefits for DHH individuals. 

This study aims to examine the long-term impact of repeated songwriting on DHH individuals’ emotional well-being and to investigate how GenAI technologies can support such ongoing practices. \revised{We focus on DHH users who have residual hearing through assistive devices, are accustomed to written-language digital communication, and self-identify as Deaf within a predominantly hearing society. This subgroup was selected to evaluate our early-stage conversational, text-based GenAI songwriting system, in line with demographic trends showing an increasing number of DHH individuals in Korea who rely on text-based communication  \cite{shin2022development,lee2023situational,cho2021experiences}.}

\begin{figure*}[ht]
  \centering
  \includegraphics[width=1\textwidth]{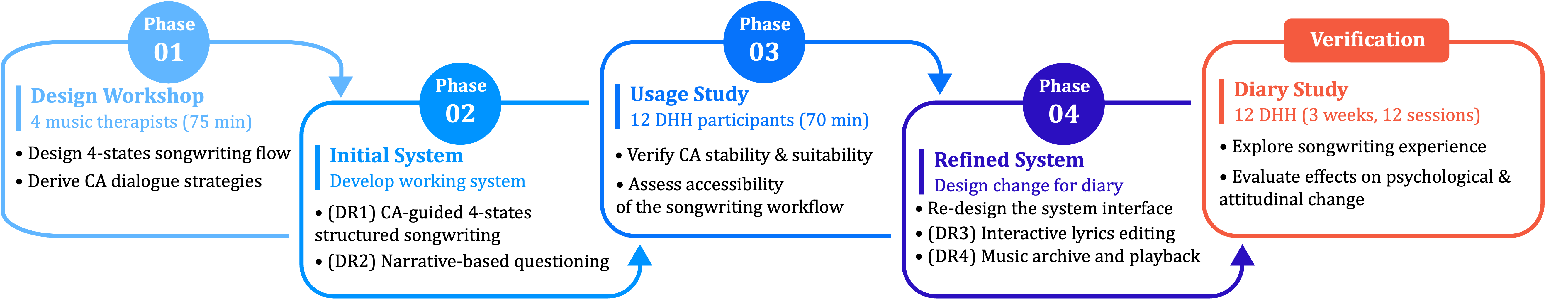}
  \caption{Overview of the research design and procedures.}
  \label{fig: process}
\end{figure*}

To achieve this goal, we propose \textit{SoulNote}, a conversational songwriting support system that enables DHH individuals to express and document emotions and daily experiences through music \cite{Youjin2025CHI, ChungHa2025}. \textit{SoulNote} aims to support sustained psychological well-being through everyday self-expression. As shown in Figure~\ref{fig: process}, we employed a user-centered design process to ensure safe and effective songwriting experiences for DHH users \cite{Abras2004}. Through a design workshop with four music therapists, \revised{we identified conversational strategies and flows that foster safe self-expression and considered potential emotional risks.} A preliminary study with 12 DHH participants assessed the safety and \revised{usability} of the conversational agent (CA). We then redesigned the system into a diary-like platform and conducted a three-week, in-the-wild diary study with 12 additional DHH participants to examine emotional and behavioral changes.

Analysis of the diary study highlighted several outcomes that underscore the value of long-term engagement. The CA’s imagery-based questions helped participants visualize experiences across past, present, and future, enabling them to recognize personal values, needs, and emotions in challenging situations. The CA-generated lyrics helped participants adopt a third-person perspective on their experiences, promoting objectivity and self-acceptance. Music style recommendations contextualized to each dialogue supported participants with limited musical experience in exploring diverse emotional expressions. Over time, the accumulation of musical artifacts served as a personalized emotional resource. \revised{Participants reported shifts in their everyday attitudes, which were reflected in behaviors such as initiating conversations about emotions with family or colleagues and adopting more active coping strategies during difficult moments.} These findings suggest that iterative songwriting can cultivate lasting emotional awareness, coping strategies, and positive behavioral shifts. The contributions of this study are as follows: 

The contributions of this study are as follows:
\begin{itemize}
    \item \textbf{Multi-phase, user-centered system design.} We employed an iterative design process that involved music therapists, a preliminary study with DHH users, and a subsequent diary study. This approach produced a system that improves the feasibility and sustainability of songwriting for DHH individuals.
    \item \textbf{System development.} We developed \textit{SoulNote}, a journaling-oriented system that integrates \revised{a reflective conversational agent, an interface for lyric-editing and multimodal appreciation, and an archival library of accumulated songs to support ongoing and self-directed emotional expression.}
    \item \textbf{In-the-wild validation of long-term effects.} \revised{Through a 12-session diary study, we qualitatively identified stepwise changes in emotional self-insight, regulation strategies, attitudinal and behavioral changes in emotional disclosure and motivation when facing difficulties. This work provides early, in-situ insights into how GenAI-supported, repeated songwriting can intersect with the emotional lives of DHH individuals.}

\end{itemize}

\section{Background}
\subsection{The Psychological Effect of Songwriting on DHH Individuals}

Barriers to verbal communication in an auditory-centered society can pose challenges for DHH individuals in expressing emotions, potentially contributing to social isolation, stigma, and difficulties with self-identity and self-esteem \cite{Davidsson2018, Punch2004, Dammeyer2009, Olsson2021}. These psychosocial challenges have been shown to negatively affect their overall life satisfaction compared to hearing individuals \cite{Vernon1999, Altshuler1962, Remvic1969}. Against this backdrop, songwriting has gained attention as a potential means of emotional recovery that embraces sensory differences \cite{Small1998}. Prior studies have highlighted that songwriting, which involves structuring and expressing one’s feelings and experiences through the symbolic languages of lyrics and music, can promote self-expression, self-efficacy, and social connectedness among DHH individuals \cite{HyeonBeom2025, Ward2016, soderberg2016}. In line with these potentials, researchers have developed songwriting support systems that deliver music through visual and tactile feedback \cite{ChungHa2025, Youjin2025CHI, Lin2022, soderberg2016}. As an example, Lee \textit{et al.} \cite{ChungHa2025} \revised{introduced a multimodal assistive tool that maps musical structure and performance to visual and vibrotactile cues, enabling DHH users to explore new expressive modalities and demonstrating the feasibility of songwriting without relying on audition.} These systems have expanded opportunities for musical self-expression and emotional release, while also showing potential to positively influence attitudes toward music and perceptions of deaf identity \cite{ChungHa2025, Youjin2025CHI, soderberg2016}. In addition, technical approaches to sign-language–based music that reflect DHH cultural identity have provided opportunities to share musical culture with hearing communities \cite{soderberg2016, Youjin2025}. 

More recently, research has begun to explore how music GenAI may enhance DHH individuals’ sense of agency in sound experiences by enabling autonomous music creation \cite{Youjin2025CHI}. Even with the support of music GenAI, however, DHH individuals may still struggle to articulate their emotions as lyrics or to translate them into musical forms that accurately capture their intended expression without interactive prompting during the songwriting process. As these challenges may limit sustained engagement, prior studies have often focused on the short-term emotional outcomes of single-session songwriting \cite{Youjin2025, ChungHa2025, Youjin2025CHI, soderberg2016}. As a result, research has yet to explore the deeper emotional and behavioral changes that may emerge from repeated songwriting practices.

\subsection{The Effects of Repetitive Self-expression}
Repeatedly expressing and recording feelings and thoughts plays an important role in mental health \cite{James1997, Pennebaker2011, Allison2011}. Journaling, including diary and reflective writing, helps individuals externalize both positive and negative daily experiences, which can facilitate emotional release, regulation, and reflection \cite{Riddell2020, Cheryl2011, Ullrich2002, Karen2012}. \revised{Psychological work on affect labeling indicates that putting feelings into words can down-regulate negative affect and support emotion regulation, which suggests one mechanism by which journaling produces benefits \cite{lieberman2007putting, torres2018}.} Regular expressive practices may also reduce suppressed emotions and strengthen self-regulation by clarifying specific experiences  \cite{Lepore2002, James1993}. From this perspective, songwriting can be regarded as analogous to journaling, as an expressive practice that externalizes inner experiences and allows emotions to be repeatedly released and regulated. Songwriting has been used in educational and therapeutic settings to address emotional difficulties across populations, typically through multi-session rather than single-session interventions  \cite{Felicity2019, Dalton2006, Eickholt2022}. \revised{Repeatedly naming and structuring emotions through lyrics and musical elements can be particularly helpful for people who struggle to identify and describe feelings, such as those with alexithymia or low emotional granularity, by providing an accessible medium that gradually enhances emotional clarity and differentiation \cite{baker2018meaning}.} Repeated songwriting has been associated with promoting student self-reflection \cite{Felicity2011}, enhancing emotional expression and communication \cite{McFerran2006, Callaghan2009, Sheri2003, Grocke2009, Dalton2006}, increasing treatment motivation in individuals with substance abuse disorders \cite{Silverman2012}, and reducing depressive symptoms in clinical populations \cite{Eickholt2022, Gee2019}. For instance, Baker \textit{et al.} \cite{Felicity2019} reported that a 12-session identity-focused songwriting program for patients with brain injury improved life satisfaction and reduced inhibition in emotional expression. These studies suggest that songwriting can foster profound emotional change and behavioral improvement when practiced repeatedly and over the long term, rather than as a one-time intervention. Therefore, providing DHH individuals with songwriting as a sustainable means of self-expression holds promise as a meaningful tool for fostering not only musical creativity but also emotional growth. 

\subsection{Generative AI for Enhanced Accessibility in Expression}
Supporting sustainable songwriting for DHH individuals requires a technological foundation that lowers barriers to participation. In particular, technologies are needed that enable users without musical training to express their personal narrative through music. Drawing on these prior studies, we explored technologies aimed at enhancing accessibility to lyrics (section \ref{sec: 2.3.1}) and music (section \ref{sec: 2.3.2}) creation for DHH individuals.

\subsubsection{Conversational agent.} 
\label{sec: 2.3.1}
Recent studies have shown that Large Language Model (LLM)-based CAs can provide critical support for individuals who face difficulties in verbal expression \cite{Taewan2024, Nepal2024, Woosuk2024, Eunkyung2024}. By scaffolding the process of organizing thoughts and constructing personal narratives, CAs help users articulate their experiences more effectively and explore deeper layers of self-expression. For example, \textit{MindfulDiary} \cite{Taewan2024} guided psychiatric patients in recalling and verbalizing past experiences through structured dialogue, particularly when they were unsure how to begin expressing themselves, and encouraged them to record these experiences regularly. Similarly, \textit{ExploreSelf} \cite{Inhwa2025} demonstrated how an LLM can provide reflective prompts and feedback that elicit deeper self-reflection. These findings suggest that CAs play a meaningful role in lowering the barriers to sustained self-expression in text-based journaling. Songwriting also requires individuals to organize and externalize their inner experiences through the language of lyrics and music. Accordingly, CAs can provide essential support for DHH individuals, particularly those with limited musical background, by guiding them in exploring and expressing their narrative through songwriting. 

\subsubsection{Content generative AI} 
\label{sec: 2.3.2}
Recent advances in content GenAI have increased access to multimedia creation, enabling images \cite{Changhoon2018, Sam2024}, videos \cite{Wang2024}, and music \cite{Yueyue2023, Zhang2021} to be generated from text prompts without prior artistic training. Prompt-based models allow non-experts to create diverse creative outputs quickly, lowering barriers to participation in artistic expression. In music, GenAI has proved particularly useful because composition traditionally required extensive technical knowledge \cite{Youjin2025CHI2, Louie2022}. For instance,  Louie \textit{et al.} \cite{Louie2020} introduced \textit{CoCoMo}, an AI-assisted co-creative system that lets novice users compose melodies by iteratively refining AI suggestions. In therapeutic contexts, Huang \textit{et al.} \cite{Louie2022} explored music generative models to create personalized relaxation music for stress reduction. \revised{Evaluations of these models indicate they can produce stylistically diverse music that approximates intended genres or moods \cite{ji2020comprehensive}.} Building on these developments, Choi \textit{et al.} \cite{Youjin2025CHI} \revised{proposed a multimodal songwriting system that integrates music GenAI, allowing DHH individuals to generate songs from text descriptions of desired musical elements.} The growing interest in inclusive music creation highlights GenAI’s potential as a foundation for DHH songwriting. 

Combining large language model-based conversational agents with music GenAI makes it possible to design journaling-oriented songwriting systems in which DHH users can explore emotions through dialogue, express them in music, and record them repeatedly.

\section{Ethical Considerations}
All studies were approved by the Institutional Review Board (IRB) of [anonymity] University in Korea. Before participation, all participants were provided with detailed information about the study procedures and related activities, asked to complete a pre-study survey, informed about the use of the system, and guided through an ethical consent process for the interviews. 

Although the participant groups did not specifically include individuals diagnosed with clinical mental illness, such as depression, we considered the possibility that \textit{SoulNote} could elicit sensitive or potentially traumatic personal expressions. To address this, we adopted the following ethical safeguards. First, to minimize ethical risks during interactions with the CA, professional therapists were actively involved throughout the system’s design and validation. They continuously reviewed potential emotional risks and guided the study design and execution. During the lab-based usage study, a therapist monitored participants in real-time, and post-session analyses of conversation logs were conducted to further review potential risks before proceeding to the in-the-wild diary study. Second, for the studies involving DHH participants, we conducted pre-surveys to measure depression \cite{Kroenke2001}. In collaboration with professional therapists, we screened for high-risk individuals during recruitment to ensure participants’ eligibility and safety. Finally, during the three-week diary study, participants were encouraged to take breaks at any time by notifying the researchers through a real-time messaging channel. The first author closely monitored interactions with \textit{SoulNote} for potential risks, including mentions of self-harm or harm to others, indications of heightened distress, or emotional triggering from LLM-generated content. \revised{These were defined as adverse events. Across 12 participants and up to 12 sessions each, no adverse events were observed, and no participants withdrew because of emotional distress.} 

\section{Design and Development of SoulNote}
\subsection{User-centered Design Process}
To develop a system that enables DHH users to independently and safely participate in songwriting anytime and anywhere through GenAI while promoting emotional expression, we adopted a multi-phase user-centered design methodology (Figure~\ref{fig: process}) \cite{Abras2004}. First, we collaborated with four music therapists specializing in DHH clients to design the CA-led songwriting process and conversational strategies (Phase 1). Subsequently, we developed an initial system based on this design (Phase 2). We then conducted a lab-based usage study with 12 DHH users to examine the stability of the system and the appropriateness of the songwriting procedure (Phase 3). Finally, we refined the prototype to improve usability by considering diary-style usage scenarios, which resulted in the system’s final version (Phase 4).

\subsubsection{Design workshop (Phases 1 and 2).} 
To enhance the emotional impact of songwriting, we conducted a 70-minute online design workshop with four music therapists specializing in music therapy for DHH individuals (Table~\ref{tab: psychotherapists}). The workshop consisted of individual design sessions and group discussions. Several key themes were identified and systematically incorporated into both the CA’s flow and the design of its conversational strategies (see Appendix~\ref{adx: Design workshop} for details).

\begin{itemize}
    \item {\textit{A challenge in songwriting without structured guidance (theme 1):}} The therapists anticipated that DHH individuals, who often have fewer opportunities for self-expression and access to musical experiences, might struggle with songwriting without guidance. Accordingly, they emphasized the need to design a structured songwriting process grounded in music therapy.
    \item{\textit{Inaccessibility of technical musical questioning in dialogue-based songwriting (theme 2):}} Therapists noted that general or technical questions about musical elements can be difficult for DHH participants to answer. To elicit more concrete musical material, they proposed using narrative-based questioning (e.g., \textit{memories, imagery, daily experiences}) and translating the responses into musical parameters during the dialogue.
\end{itemize}

\begin{table*}[t]
\small
\begin{center}
  \caption{Demographic information of the music psychotherapists in the design workshop}
  \label{tab: psychotherapists}
    \begin{tabularx}{\linewidth}{c|c|c|Y|Y|Y}
    \toprule
    ID & Gender & Age & Job title & Specific field & \makecell{Years of experience} \\
    \midrule
    T1 & Female & 40 & Clinical music therapist & Percept (Psycho) & 11  \\ \hline
    T2 & Female & 32 & Licensed music therapist & Percept (Psycho) & 5  \\ \hline
    T3 & Female & 35 & Licensed music therapist & Percept (Psycho) & 6   \\ \hline
    T4 & Female & 36 & Licensed music therapist & Percept (Psycho) & 10 \\
    \bottomrule
    \multicolumn{6}{>{\raggedright\arraybackslash}p{\dimexpr\linewidth-2\tabcolsep\relax}}{%
    \small Job title is defined as a licensed music therapist (Licensed music therapist) who also holds a doctoral degree in clinical music therapy (Clinical music therapy). The specific field indicates the participant’s primary expertise in music perception (Percept) and their supplementary experience in music psychotherapy (Psycho).}
\end{tabularx}
 \vspace{-10pt}
\end{center}
\end{table*}

\subsubsection{Lab-based usage study (Phases 3 and 4).} 
\revised{To evaluate the usability and stability of the initial prototype developed from the design workshop, we conducted a 70-minute usage study with 12 DHH participants who had residual hearing and regularly used written Korean for digital communication (Table~\ref{tab: participants}: Usage study)}. Each participant completed a songwriting activity using our system, followed by an individual interview about their experience. A music therapist joined all sessions to monitor the process. Additional details of the process and results are provided in the Appendix~\ref{adx: usage study}. The key results are listed below.
\begin{itemize}
    \item {\textit{The potential of a safe and user-friendly system:}} The music therapist’s notes indicated that the structured songwriting procedure and CA questions offered a protective scaffold that helped prevent emotional over-immersion and ensured that sessions concluded appropriately. In post-session interviews, participants reported that they could complete the songwriting task without major difficulty. Several participants also described long-standing challenges expressing their feelings in everyday conversations. In contrast, they found \textit{SoulNote} to be a supportive medium for articulating thoughts and emotions through music in collaboration with the CA. These responses suggest that the system offers a promising and accessible avenue for self-expression.
\end{itemize}
However, several improvements needed for long-term deployment were identified.
\begin{itemize}
    \item{\textit{Difficulty in lyrics revision through dialogue (theme 3):}} All participants actively engaged in lyrics creation: some requested specific revisions from the CA (n=5), while others copied generated text into a notepad for direct editing (n=7). However, they reported difficulty with the linear, dialogue-based interaction for lyrics revision, highlighting the need for a more accessible method for editing lyrics. 
    \item{\textit{Need for retrospective reflection through re-listening to songs (theme 4):}} Most participants expressed a strong desire to save the generated songs for repeated listening. They perceived these songs as deeply personal expressions that offered comfort, encouragement, and a reflection of their individuality, distinct from commercially produced music.
\end{itemize}

\begin{table*}[t]
\small
\begin{center}
\caption{Demographic information of the DHH participants in the usage and diary study.}
  \label{tab: participants}
    \begin{tabularx}{\linewidth}{c|c|c|c|c|c|c|c|c|c}
    \toprule
    Study & ID & Gender & Age & Hearing loss & \makecell{Hearing loss age} & \makecell{Aid\\device} & \makecell{\revised{Communication}\\\revised{method}} & \makecell{Knowledge\\of GenAI} & \makecell{PHQ-9\\score} \\
    \midrule
    \multirow{12}{*}{\makecell{Usage\\Study}} & U1 & F & 54 & Mild & Infant & CI + HA & W/T + KSL & No Knowledge & Minimal \\
    & U2 & F & 43 & Mild & Unknown & CI & W/T & Knowledge & Minimal \\
    & U3 & F & 47 & Mild & Infant & CI & W/T & Knowledge & Mild \\
    & U4 & F & 53 & Mild & Infant & CI + HA & W/T & No knowledge & Mild \\
    & U5 & F & 46 & Profound & Infant & CI & W/T + KSL & No knowledge & Minimal \\
    & U6 & F & 49 & Moderate severe & Innate & CI + HA & W/T &Knowledge & Minimal \\
    & U7 & F & 32 & Moderate severe & Infant & CI & W/T & Experienced & Mild \\
    & U8 & F & 53 & Profound & Infant & CI & W/T & No knowledge & Mild \\
    & U9 & M & 28 & Severe & Innate & CI & W/T + KSL & Experienced & Moderate \\
    & U10 & M & 22 & Severe & Innate & CI & W/T + KSL & Experienced & Mild \\
    & U11 & M & 32 & Moderate severe & Unknown & CI & W/T & Experienced & Mild \\
    & U12 & M & 32 & Moderate severe & Innate & CI + HA & W/T &Experienced & Moderate \\
    \hline
    \multirow{12}{*}{\makecell{Diary\\Study}} & P1 & F & 46 & Moderate severe & Infant & CI + HA & W/T & Experienced & Mild \\
    & P2 & F & 55 & Mild & Infant & CI & W/T & Experienced & Mild \\
    & P3 & F & 28 & Moderate severe & Infant & CI & W/T + KSL & Knowledge & Minimal \\
    & P4 & M & 28 & Mild & Innate & CI & W/T & Knowledge & Mild \\
    & P5 & M & 67 & Moderate & Infant & CI + HA & W/T & Experienced & Mild \\
    & P6 & F & 53 & Mild & Infant & CI + HA & W/T & Knowledge & Minimal \\
    & P7 & F & 39 & Moderate severe & Innate & CI & W/T & Experienced & Mild \\
    & P8 & M & 27 & Mild & Innate & CI & W/T + KSL & Experienced & Minimal \\
    & P9 & M & 28 & Mild & Innate & CI & W/T + KSL & Experienced & Moderate \\
    & P10 & F & 26 & Moderate severe & Infant & HA & W/T + KSL & Experienced & Mild \\
    & P11 & F & 32 & Moderate severe & Innate & CI + HA & W/T + KSL & Experienced & Minimal \\
    & P12 & M & 30 & Mild & Infant & HA & W/T + KSL & Experienced & Minimal \\
    \hline
    \multirow{10}{*}{\makecell{\revised{Additional}\\\revised{Study}}} & A1 & F & 30 & Mild & Infant & CI + HA & W/T + KSL & Knowledge & Minimal \\
    & A2 & F & 38 & Moderate & Infant & CI & W/T + KSL & Experienced & Minimal \\
    & A3 & F & 25 & Moderate severe & Unknown & CI & W/T + KSL & Experienced & Minimal \\
    & A4 & F & 46 & Mild & Infant & CI + HA & W/T + KSL & Experienced & Minimal \\
    & A5 & F & 30 & Moderate & Innate & CI & W/T + KSL & Experienced & Mild \\
    & A6 & M & 28 & Moderate severe & Innate & CI & W/T + KSL & Knowledge & Minimal \\
    & A7 & F & 52 & Moderate severe & Infant & CI + HA & W/T & Experienced & Minimal \\
    & A8 & M & 25 & Mild & Innate & CI & W/T & Knowledge & Minimal \\
    & A9 & M & 30 & Mild & Unknown & CI + HA & W/T & Knowledge & Minimal \\
    & A10 & F & 54 & Mild & Infant & CA & W/T & Knowledge & Minimal 
    \\
    \bottomrule
    \multicolumn{10}{>{\raggedright\arraybackslash}p{\dimexpr\linewidth-2\tabcolsep\relax}}{%
    \small Hearing loss (with aid device): Mild (20–34 dB), Moderate (35–49 dB), Moderate severe (50–65 dB), Severe (65–79 dB), and Profound (80–94 dB) with aid device; Aid device: Cochlear Implant (CI) and Hearing Aid (HA); \revised{Communication method: Korean Sign Language (KSL), Writing and Typing (W/T);} Knowledge of GenAI: prior usage (Experienced), awareness only (Knowledge), and no knowledge of GenAI (No knowledge); PHQ-9 scores were categorized as follows: Minimal (0–4), Mild (5–9), Moderate (10–14), Moderately severe (15–19), and Severe (20–27). From the moderately severe category ($\geq 15$), evidence-based guidelines recommend active intervention, including pharmacotherapy or psychotherapy.}
\end{tabularx}
 \vspace{-10pt}
\end{center}
\end{table*}

\subsection{Designing \textit{SoulNote} for Sustainable Songwriting}
\begin{figure*}[ht]
  \centering
  \includegraphics[width=1\textwidth]{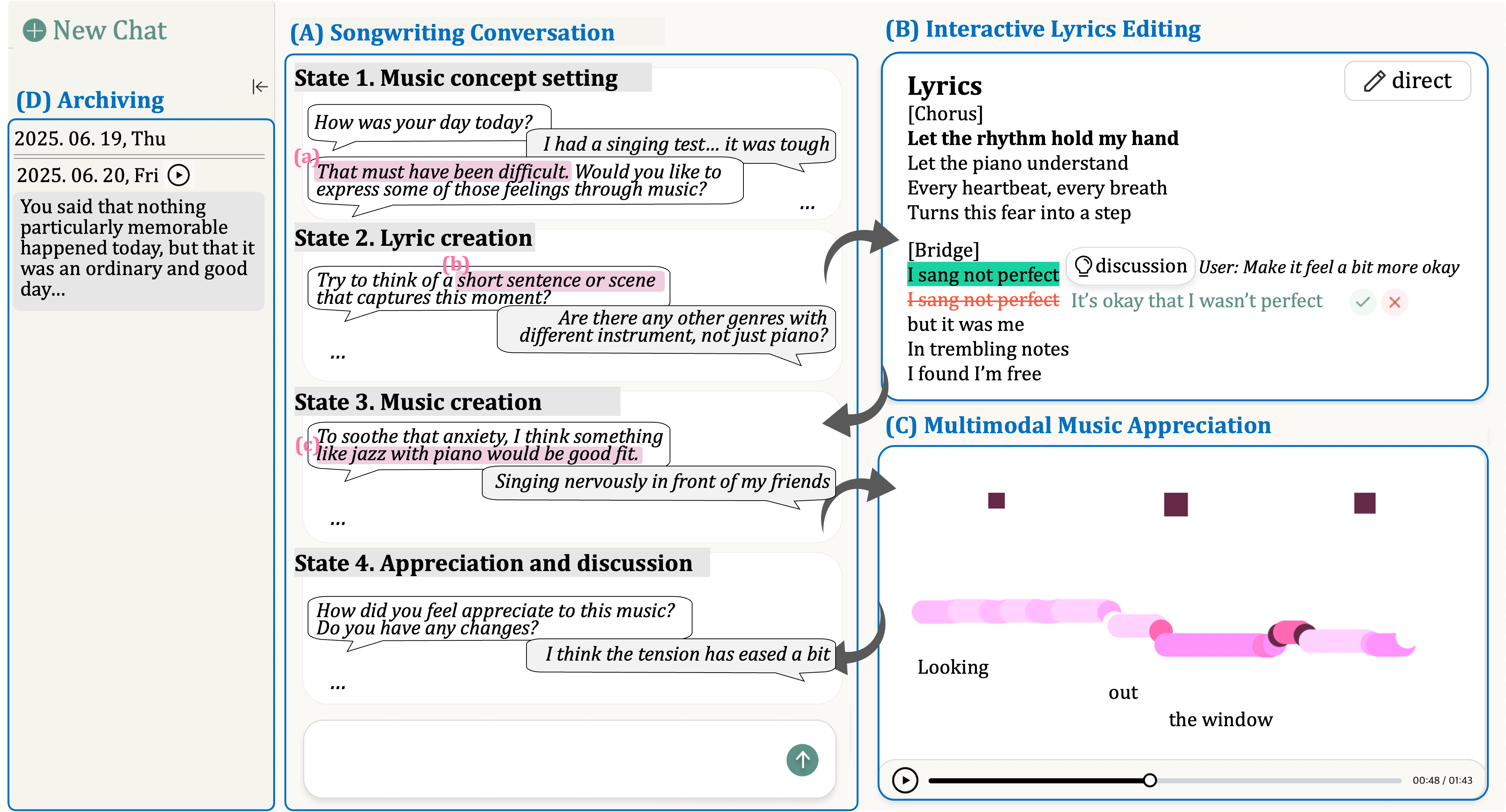}
  \caption{Interfaces and features of \textit{SoulNote}. (A) Conversation interface for songwriting with four dialogue phases (\textit{highlights indicate CA strategies such as supportive and unembellished response (a), imagery-based questioning (b), and context-based music style recommendation (c)}), (B) Interactive lyrics editing interface for modifying CA-generated lyrics, (C) Music appreciation interface with visual assistance, and (D) interface for archiving and reviewing previous songwriting results, including conversation summaries and music replay.}
  \label{fig: interface}
\end{figure*}

Table~\ref{tab: designelement} summarizes the design requirements (DRs) derived from two preliminary studies that guided the design of the \textit{SoulNote} system, along with the corresponding key features implemented to support its long-term deployment. The design workshop focused on the CA-related design of the songwriting process, while the lab-based usage study addressed \revised{practical use and usability considerations} for delivering the system in a diary format. Based on these findings, \textit{SoulNote} was designed around four core interfaces according to DRs (Figure~\ref{fig: interface}): (a) a conversation interface for songwriting, (b) a lyrics editing interface, (c) a multimodal music appreciation interface, and (d) an interface for summarizing and recording previous songwriting outcomes. 

\begin{table*}[ht]
\small
\begin{center}
  \caption{Summary of preliminary study findings and the design elements incorporated into \textit{SoulNote}}
  \label{tab: designelement}
    \begin{tabularx}{\linewidth}{p{1.3cm}|p{1.4cm}|X|X|X}
    \toprule
    Study & System & Theme & Design Requirement & Feature \\
    \midrule
    \multirow{2}{*}{\makecell[l]{Design\\Workshop}} & \multirow{2}{*}{\makecell[l]{CA/\\prompting}} & Theme 1. A challenge in songwriting without structured guidance & DR1. Structured songwriting guided by CA & Four-state songwriting flow \\ \cline{3-5}
    & & Theme 2. Inaccessibility of musical questioning in dialogue-based songwriting & DR2. Narrative-based questioning for musical expression support& \textbullet Imagery-based questions\newline \textbullet Context-based music style recommendation\newline \textbullet Supportive and unembellished response \\ \hline
    \multirow{2}{*}{\makecell[l]{Usage\\Study}} & \multirow{2}{*}{Interface} & Theme 3. Difficulty in lyrics revision through dialogue & DR3. Direct lyrics editing support & Interactive lyrics editing interface\\ \cline{3-5}
    & & Theme 4. Need for retrospective reflection through re-listening to songs & DR4. Archiving and reflection support & Music archive and playback interface \\
    \bottomrule
\end{tabularx}
 \vspace{-10pt}
\end{center}
\end{table*}

\subsubsection{\textbf{Songwriting conversation interface (Figure~\ref{fig: interface}-A)}}\hfill

\textbf{DR1. Structured songwriting guided by CA.}  For DHH participants unfamiliar with self-expression and musical activities, an unstructured songwriting process can be burdensome and challenging (theme 1). To address this challenge, we designed (1) a structured process in which the CA guides users through their songwriting, drawing on prior LLM-based studies that proposed structured and step-by-step approaches to support self-expression \cite{Skeggs2025, Liu2025, Zhou2025}, and (2) the integration of lyrics and music GenAI to \revised{lower the technical barriers to songwriting for our target subgroup of DHH users} \cite{Youjin2025CHI, Huang2024}. Ultimately, drawing on therapeutic procedures commonly applied in clinical practice \cite{Yixiao2024, Bougueng2022}, we developed a CA-guided four-state songwriting process: 

\begin{itemize}
    \item \textit{Music concept setting.} The CA supports users in defining a music concept based on their daily events, thoughts, and emotions.
    \item \textit{Lyric creation.} The CA generates draft lyrics based on the user’s narrative, and users can subsequently refine them using an interactive lyrics editing interface.
    \item \textit{Music creation.} Through dialogue, the user and the CA specify musical elements such as mood, genre, instruments, and vocals, and the CA then generates corresponding music with a GenAI model.
    \item \textit{Appreciation and discussion.} The generated music is presented with visualization, enabling the user to experience the outcome and engage in reflective discussion to conclude the session.
\end{itemize}
\textbf{DR2. Narrative-based questioning for musical expression support.} \hfill

For DHH participants unfamiliar with songwriting, eliciting responses requires focusing on their personal narratives to guide lyrics and musical expression (theme 2). In consultation with therapists, we refined three prompting strategies to ensure safe and accessible interaction. First, the CA was designed to adopt (1) a \textit{supportive dialogue without judgment or embellishment}, providing dialogue that helps users articulate their situations rather than interpreting them. This supportive dialogue strategy, grounded in music therapy practice, does not involve delving deeply into specific situations but instead provides supportive interaction mediated through music, guiding users toward musical expression \cite{Wheeler2015}. Second, acknowledging that many DHH users have limited musical expression, our design drew on two insights: the tendency of DHH individuals to appreciate music visually \cite{Daphne2006, Kyrie2024, JaeYoung2025} and the effectiveness of guided imagery in recalling concrete situations and emotions \cite{Hackmann2011, Lusebrink1990}. Based on these, we incorporated (2) a \textit{concrete imagery-based questioning strategy}, encouraging storytelling through message or image, e.g., \textit{``When you think about your current emotions?'',} or \textit{``What kind of scene comes to mind?''.} Additionally, as DHH users are often not accustomed to independently selecting musical elements that match their narratives, it was necessary to provide recommendations aligned with their musical intentions \cite{Baker2005}. To this end, we introduced (3) a \textit{context-based music style recommendation strategy}, where the CA suggests musical elements aligned with the user’s narrative, e.g., \textit{``To express your current calm and delicate emotions, the piano might be a good fit.''} These strategies were validated in the usage study, where therapist monitoring confirmed that they facilitated safe, natural self-expression without evoking deep negative emotions.

\subsubsection{\textbf{A lyrics editing interface (Figure~\ref{fig: interface}-B)}}\hfill

\textbf{DR3. Direct lyrics editing support.} For DHH participants in the usage study who valued lyrics creation, revising lyrics exclusively through a conversational interface proved insufficient. In particular, the process of specifying lines and confirming changes through multiple dialogue turns was cumbersome and reduced accessibility (theme 3). To address this, we introduced a new interface that combined conversational revision with direct editing \cite{Laban2024, Wang2024}. \textit{SoulNote} was revised to allow users to (1) directly edit lyrics themselves or (2) select specific segments for revision through CA prompts (e.g., \textit{``Make this part more metaphorical''}). Revisions were displayed instantly, and users could confirm or discard them using checkmark and X icons.

\subsubsection{\textbf{Multimodal music appreciation and archive (Figure~\ref{fig: interface}-C and D)}}\hfill

\textbf{Music appreciation.} To enhance music appreciation, \textit{SoulNote} implements a multimodal appreciation feature that visualizes lyrics and musical elements simultaneously, drawing on prior research \cite{Youjin2025CHI}. Prior work has highlighted that for DHH individuals, effective music appreciation involves not only transforming musical elements into visual forms but also visualizing lyrics. Accordingly, rhythm is represented by rectangular blocks, pitch by dot positions, lyrics by animated height variations reflecting vocal pitch, and mood/emotion by colors and fonts. This feature extends DHH users’ music experiences by leveraging music-to-visual conversion feedback, which supports perception for hearing aid users \cite{soderberg2016, HyeonBeom2025}.

\textbf{DR4. Music archive and playback.} \hfill

The songwriting process produced musical outcomes that reflected personal narratives, which DHH participants in the usage study regarded as valuable resources. They expressed a preference for re-appreciating their creations with multimodal support and engaging in reflection during the process (theme 4). Prior research has shown that summarizing and revisiting personal information supports reflection and self-understanding, whether through journaling practices or digital interfaces \cite{Sas2009, Ullrich2002}. Building on these insights, we implemented an archive and playback feature. This feature stores lyrics, music, and session summaries, including experiences, emotions, and themes, and enables users to revisit and reflect on their works with visualization through a sidebar interface.

\subsection{Implementation of \textit{SoulNote}}
\subsubsection{\textbf{Frontend and backend.}}
\textit{SoulNote} is a web-based application that integrates GenAI. It consists of two main components: (1) a frontend for user interface and visualization, and (2) a backend for dialogue processing, music generation, and analysis. The front-end interface was built with Next.js\footnote{https://nextjs.org/} and deployed via Vercel\footnote{https://vercel.com/home}, while the back end was developed with Flask\footnote{https://flask.palletsprojects.com/en/stable/} and deployed on AWS EC2\footnote{https://aws.amazon.com/ec2/}.

The CA is powered by OpenAI’s GPT-4.1\footnote{\url{https://openai.com/index/gpt-4-1/}} \revised{(temperature = 0, frequency and presence penalties = 0, and all other parameters left at their default values)} to facilitate songwriting conversations. For music generation, the system analyzes user dialogue to construct lyrics and music prompts, which are then processed through the Mureka.ai API\footnote{\url{https://www.mureka.ai/}} \revised{(model: \textit{mureka-6})}. \revised{Mureka.ai is a commercial text-to-music platform that uses deep generative models trained on large-scale music corpora to transform text prompts and lyrics into complete, production-ready songs with melody, harmony, instrumentation, and mixing.} Musical analyses required for visualization are extracted via the MusicAI API\footnote{https://music.ai/} and mapped to visual elements. All conversation logs, generated lyrics, music files, and visualization data are stored in Supabase\footnote{https://supabase.com/}, ensuring consistency across sessions and enabling archiving functionality.

\subsubsection{\textbf{Prompting design of conversational agent.}}
In alignment with the four-states CA-guided structured songwriting process—comprising music concept setting, lyric creation, music creation, and appreciation/discussion—the CA was implemented using a state–slot technique \cite{Wang2024-2}. This technique designs specific variables (slots) required at each state, collects corresponding pieces of information from user responses, and transitions the system to the next state once all required slots are filled. In consultation with music therapists, the prompts were carefully designed to elicit the variables required at each state and to draw out the elements needed for songwriting (Table ~\ref{tab: state and slot} in ~\ref{adx: prompt engineering}). For example, in the music concept state, the goal is to identify a musical theme based on daily experiences and emotions. To achieve this, slots such as \textit{today\_emotion, difficulty, expression} are extracted, and upon completion, the system proceeds to the next state. For this process, we designed the CA prompting with four components. The \textit {general prompt} defines the CA’s role, specifies the rules for questioning and responding, and provides guidance on the overall songwriting states. The \textit{state-specific prompt} generates questions to fill the slots required for each of the four songwriting states. User conversations are stored in a database, and at each turn, a \textit{slot-extraction prompt} identifies the relevant slot and determines whether the current state has been completed. Finally, the \textit{summary prompt} guides the summarization of key content from the completed songwriting for archiving. The details of which are provided in the Appendix \ref{adx: prompt engineering}. 

\subsection{Usage Scenario}
The following is an example usage scenario of Lora, a fictional user, demonstrating the interface shown in Figure~\ref{fig: interface}. At the end of her day, Lora logs into \textit{SoulNote}. When the system begins, the CA provides a summary of the key points from the previous activity. She then clicks the sidebar to revisit yesterday’s music through the multimodal appreciation interface, recalling past emotional experiences. Then, the CA initiates a new songwriting by asking, \textit{``How was your day today?''} Lora responds that she felt anxious and exhausted because of a singing test. The CA responds supportively \textit{``That must have been difficult''} and then guided the conversation further by asking whether she would like to set today’s music theme around this situation. From this dialogue, the CA identifies the day’s musical concept and guides Lora to create lyrics. When the CA asks, \textit{``Think of a short phrase or image that can represent that moment,''} she replies, \textit{``Standing nervously in front of my friends, facing their expressionless faces.''} After further discussion, the CA generates a draft of lyrics. She then edits the generated lyrics either by directly modifying the words through the lyrics editing interface and by asking the CA to make specific changes, \textit{``Make this part more metaphorical''.} Once the lyrics are finalized, the CA engages Lora in a dialogue about designing the musical elements. Since she is unsure which genre might fit, the CA suggests that calm classical music may be suitable to ease anxious feelings. Lora accepts the suggestion, chooses the genre, and adjusts musical parameters such as tempo. Then, the CA generates the music, which is presented through the appreciation interface along with visualizations. After she appreciates music, the CA asks, \textit{``How did you feel while experiencing this music?''.} Lora responds that the anxiety and stress from today's test were alleviated through the music. At the conclusion of the songwriting session, she is able to access in the sidebar her generated lyrics and music, along with a summary of the dialogue.

\section{In-the-Wild Diary Study}
\subsection{Participants}
We conducted a three-week, in-the-wild diary study to evaluate the longer-term effects of iterative songwriting with \textit{SoulNote}. Recruitment criteria were: (1) clinical hearing impairment, (2) ability to use computer-based systems and communicate with the conversational agent in written Korean, and (3) interest in musical self-expression or journaling, regardless of prior songwriting experience. The recruitment flyer described the study goal, participation period, procedures, and activities. Recruitment occurred over approximately two weeks through hard-of-hearing communities and hearing care centers \revised{across multiple regions of Korea to avoid concentrating the sample in a single local community.} We administered a pre-survey to promote diversity and screen for depression [41]. \revised{We confirmed that no respondents exceeded our predefined high-risk threshold, so no one was excluded for safety reasons.} Twelve participants were enrolled, aged from their 20s to their 50s (mean = 38.25, SD = 13.17), with hearing levels ranging from mild to profound (Table~\ref{tab: participants}: Diary study). \revised{All participants communicated in written Korean; several also reported using Korean Sign Language in daily life.} Each participant received a stipend of approximately USD 200. 

\subsection{Process and Materials}
\revised{To examine the initial feasibility of deploying SoulNote in real-world settings, we used a single-condition longitudinal design to provide a comfortable, low-burden participant experience \cite{Taewan2024, Inhwa2025}}. DHH participants used the system freely over three weeks. Drawing on prior work indicating that approximately ten sessions are needed to observe emotional or behavioral change \cite{Chen2021}, we asked participants to complete twelve songwriting sessions, aiming for roughly four sessions per week. Before beginning the study, all participants completed a three-day video tutorial and conducted at least one practice session to familiarize themselves with the system. During the three-week study, each participant completed the 12 required sessions. After every session, participants rated their satisfaction with the process (\textit{``How satisfied are you with today’s music creation process with the CA?''}) and with the musical outcomes (\textit{``How satisfied are you with the lyrics and music created today?''}) on a 7-point Likert scale \cite{Youjin2025CHI}. They also rated the perceived effort required (\textit{``How much effort did you put into today’s music creation with the CA?''}). 

\subsection{Data Analysis}
We conducted a qualitative analysis of conversational data, weekly reports, and interview transcripts, \revised{focusing on exploratory interpretation and intentionally avoiding strong claims about the system’s effects, given the small sample size and lack of comparison conditions} \cite{maxwell2010using, Virginia2006}. First, the first author and a co-author with expertise in music therapy conducted an initial thematic analysis of participants’ conversations with the CA. They identified four categories used to label the data: songwriting topic, creative intent, lyric strategy, and musical style. Next, the weekly journals and interview transcripts were reviewed to develop an initial codebook. Discrepancies were discussed and incorporated into a refined codebook. Next, codes and data were reviewed with an additional co-author who was not involved in the interviews, resulting in a final thematic structure comprising four themes (used as subsection headings) and six subthemes (used as subsubsection headings). The first theme characterizes participants’ iterative engagement with songwriting across sessions (Section~\ref{sec:6.1}). Guided by prior work on emotional development and trait emotional intelligence \cite{saarni1999development, petrides2001trait}, we interpret the remaining three themes—self-insight (Section~\ref{sec:6.2}), emotion regulation (Section~\ref{sec:6.3}), and attitudinal change (Section~\ref{sec:6.4})—as reflecting a three-dimensional emotional growth lens. Table~\ref{tab:codebook} in Appendix~\ref{appendix:D} presents the detailed codes.

\begin{figure*}[ht]
  \centering
  \includegraphics[width=1\textwidth]{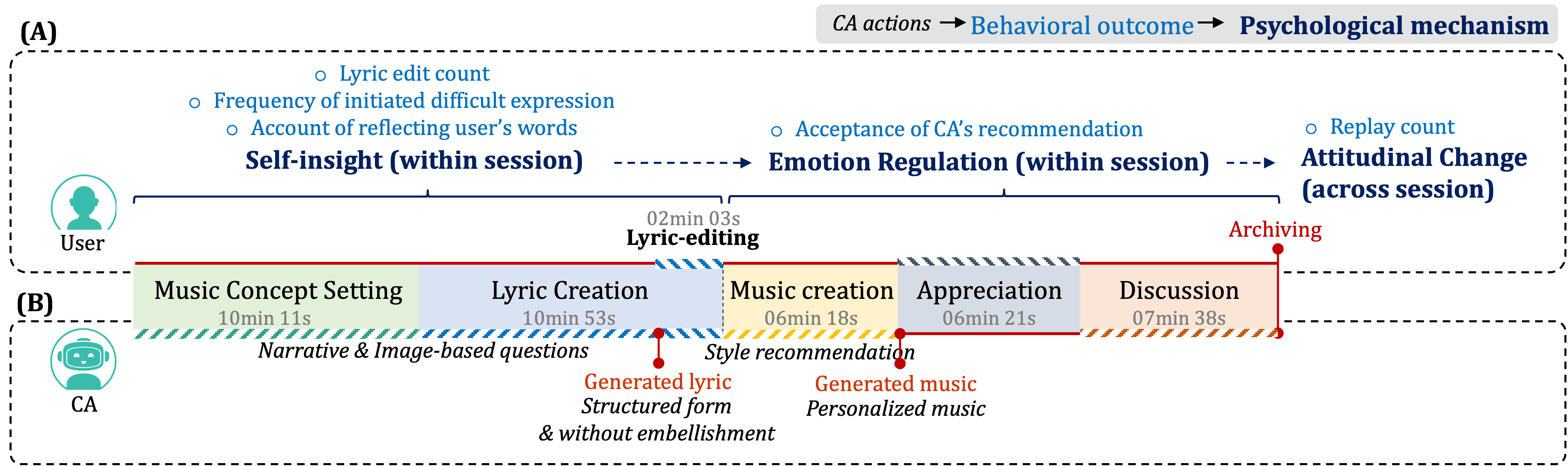}
  \caption{\revised{Overall process and theory of change in CA-assisted songwriting. The upper layer (A) represents user-side psychological processes associated with interacting with the CA,  organized into three dimensions: developing self-insight, experimenting with emotion regulation strategies, and shifts in everyday attitudes and behaviors. To complement these qualitatively derived mechanisms, we also analyzed behavioral log indicators of system engagement (\textit{e.g.,}lyric editing time and counts, frequency of initiating difficult topics, contextual music recommendations requested, and replay counts of archived songs; see Section 6). The lower layer (B) illustrates the structure of a typical session with the CA, detailing the agent’s actions (lyric generation, music generation, and archiving) and the CA’s conversational strategies at each step of the process (e.g., narrative and image-based questions). The middle timeline summarizes SoulNote’s overall workflow and the average time users spent in each task phase within a session, including music concept setting, lyric and music creation, appreciation, and discussion.}} 
  \label{fig: theoryofchange}
\end{figure*}

\revised{To complement the observational and self-report findings focused on users’ change, we additionally collected behavioral log data—per-session lyric editing time and counts, the frequency with which participants initiated difficult emotional topics, the number of contextual music recommendations requested, and replay counts of archived songs—used not as direct proxies for psychological change but as supplementary indicators of how participants engaged with the system over time (Figure~\ref{fig: theoryofchange}-A). These indicators were analyzed descriptively to triangulate qualitative themes related to self-insight, emotion regulation, and shifts in everyday attitudes, rather than to support causal claims.}


\section{Findings}
\subsection{\textbf{\revised{Iterative Songwriting as a Channel for Structuring and Expressing}}}
\label{sec:6.1}
Across 12 sessions, all participants completed the expression of their situations and emotions through lyrics and music using  \textit{SoulNote}. \revised{No instance of task failure was found,} and reported high satisfaction with both the songwriting process (M=6.04, SD=0.91) and the musical outcomes (M=6.07, SD=1.01)\revised{, and the required effort was low (M=3.66, SD=1.60)}. The average session duration was 41 minutes 24 seconds (SD=12m 56s)\revised{, as shown in Figure~\ref{fig: theoryofchange}}. 

\begin{figure*}[ht]
  \centering
  \includegraphics[width=1\textwidth]{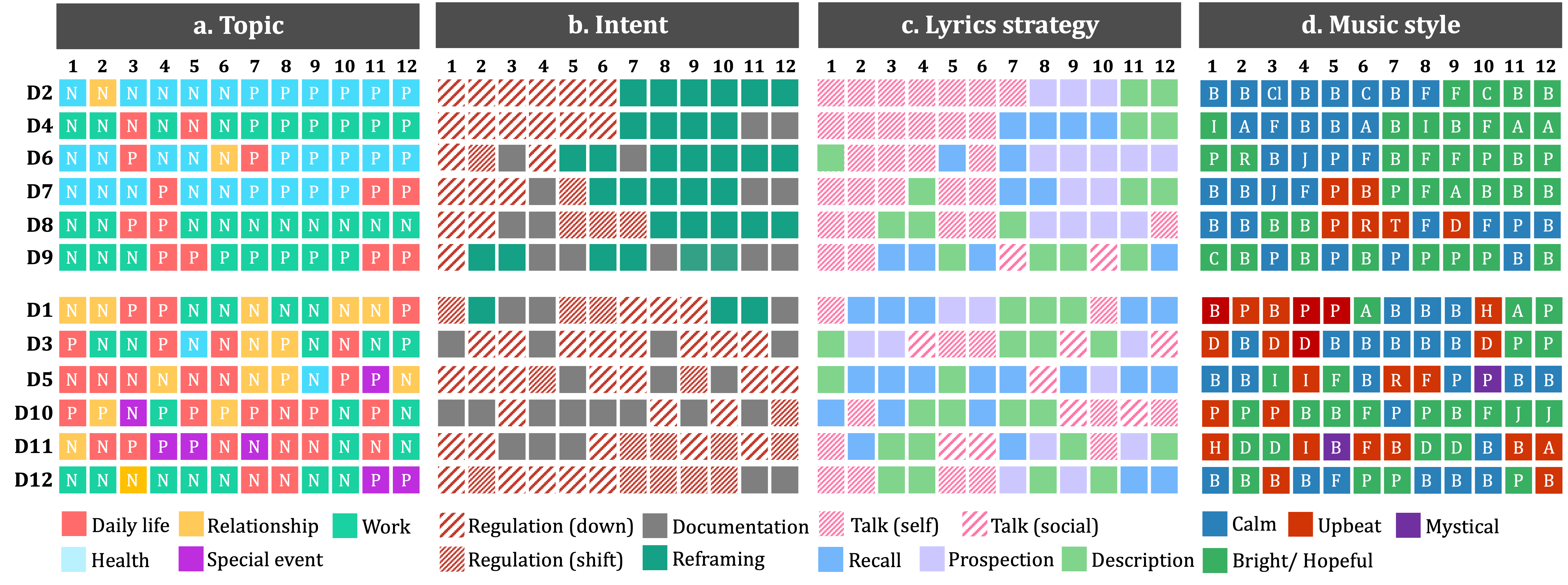}
  \caption{Individual songwriting patterns of DHH participants across 12 sessions. The patterns are categorized into five topics, four intents, five lyrics creation strategies, and four musical moods. Topics are annotated with (N) for negative and (P) for positive emotions. Music styles are represented as follows: Ballad (B), Pop (P), Acoustic (A), Folk (F), Classical (CI), CCM (C), Dance (D), Hip-hop (H), Indie (I), Rock (R), and Trot (T).}
  \label{fig: usage pattern}
\end{figure*}

Participants structured their emotions and thoughts into lyrics and music in relation to specific experiences across sessions. In practice, they typically composed music on days when they wished to record specific events or emotions. Table~\ref {tab: conversationcategory} summarizes the categories of participants’ music topics, intentions, strategies, and music style. The music topic was generally categorized into \textit{daily life}, \textit{relationships}, \textit{work}, \textit{health}, and \textit{special events}. As shown in Figure~\ref{fig: usage pattern}, participants who repeatedly mentioned the same topics primarily created music with the intention of alleviating or shifting negative emotions (\textit{regulation}) or reinterpreting difficult situations from new perspectives (\textit{reframing}). Their musical outputs were most often calm or bright/hopeful in style. In contrast, participants who brought up diverse topics were more likely to focus on emotion regulation, producing music across a wider variety of moods and genres, including ballad, pop, hip-hop, classical, and CCM, with moods that ranged from \textit{calm}, to \textit{bright/hopeful}, to \textit{upbeat}, or \textit{mystical}.

All participants liked songwriting as a way of recording positive experiences for future recollection (\textit{documentation}). To support these musical concepts and intentions, participants collaborated with the CA using a range of lyrics creation strategies, including recalling past experiences (\textit{recall}), imagining future possibilities (\textit{prospection}), directly describing present situations (\textit{description}), and composing messages addressed to themselves or others (\textit{talk}).

\begin{table*}[t]
\small
\begin{center}
  \caption{Categorization of participants' songwriting processes}
  \label{tab: conversationcategory}
    \begin{tabularx}{\linewidth}{p{1.2cm}|p{2.3cm}|X|X}
    \toprule
     & Category & Define & Sub-category \\
    \midrule
    \multirow{1}{*}{Topic} & Daily life & Themes related to everyday activities, such as rest or watching TV & Rest, watching TV, diet, meeting friends, meal, love, prayer, walking, weather 
    \\\cline{2-4}
    & Relationship & Interpersonal themes such as conflicts or longing for family, friends, or colleagues & Colleagues, family, friends, unknown\\\cline{2-4}
    & Work & Themes related to preparing for work or workplace life & Overload, job preparation, new challenge, workplace life \\ \cline{2-4}
    & Health & Themes related to health, such as auditory surgery and rehabilitation & Auditory rehabilitation, physical pain, and surgery \\ \cline{2-4}
    & Special event & Unusual or significant events, such as a friend’s wedding, are not part of the daily routine & Travel, moving, wedding, memorial day, picnic \\ 
    \hline
    \multirow{1}{*}{Intent} & Emotion \newline regulation- down & \multicolumn{2}{p{13cm}}{%
    \raggedright Intent to relieve emotions (e.g., \textit{``Today’s auditory rehabilitation was so difficult that I want to be calmed and comforted''(P2),} \textit{``I felt very anxious while studying today, and I want to reduce my anxiety''(P8)})}\\ \cline{2-4}
    & Emotion \newline regulation-shift & \multicolumn{2}{p{10.5cm}}{%
    \raggedright Intent to shift emotions (e.g., \textit{``I felt so bored today that I want to gain some energy'' (P1).})}\\ \cline{2-4}
    & Documentantation & \multicolumn{2}{p{13cm}}{%
    \raggedright Intent to document current situations and emotions in music for oneself or to share with others (e.g., \textit{``I want to preserve today’s peaceful day as music so I can listen to it later'' (P10),} \textit{``I want to record my feelings in music and share them with my colleagues'' (P11)})}\\ \cline{2-4}    
    & Reframing & \multicolumn{2}{p{13cm}}{%
    \raggedright Intent to reinterpret emotions or situations
    (e.g., \textit{``I want to express the thought that everything I am going through now will eventually pass and is necessary'' (P6),} \textit{``Through a song about good memories with my younger sister, I want to reconsider our relationship'' (P1)})}\\
    \hline
    \multirow{1}{*}{Strategy} & Talk (self) & Expressions of messages one wishes to hear to provide oneself with comfort, encouragement, or hope & Encouragement, recovery, strength, acceptance, reflection  \\ \cline{2-4}
    & Talk (social) & Expressions of messages intended to be shared with friends, colleagues, or the community & Colleagues, family, community \\ \cline{2-4}
    & Recall & Expressions based on recalling past positive or negative experiences & Colleagues, family, friends, negative experience, youth, childhood \\ \cline{2-4}
    & Prospection & Expressions based on imagining future events, such as work or life situations & Working, health, weather, travel/rest \\ \cline{2-4}
    & Description & Direct expressions that describe emotions and situations as they are & Emotion, situation\\
    \hline
    \multirow{1.5}{1.2cm}{Music Style} & Calm/Soothing & \multicolumn{2}{p{10.5cm}}{Acoustic ballad, ballad, classic, folk, pop, jazz, korean trot, nature ambient}\\ \cline{2-4}
    & Hopeful/Bright & \multicolumn{2}{p{13cm}}{Acoustic ballad, acoustic, dance, pop, folk, indie, ballad, jazz}\\ \cline{2-4}
    & Upbeat & \multicolumn{2}{p{13cm}}{Acoustic ballad, acoustic,  ballad, indie, pop, rock, hip-hop, jazz, new age}\\ \cline{2-4}
    & Mysterial & \multicolumn{2}{p{13cm}}{CCM, folk, indie, pop, rock, ballad}\\ 
    \bottomrule
\end{tabularx}
 \vspace{-10pt}
\end{center}
\end{table*}

\subsection{Cultivating Self-insight through Repetitive Lyrics Creation that Addresses Negative Experiences}
\label{sec:6.2}
Participants generally reported that the processes of discussion, generation, and revision during lyrics creation with the CA helped address their negative situations and emotions. Due to communication barriers, Many participants (n=8) had seldom been able to express their thoughts and feelings directly to others. In our study, the narrative-centered dialogue for lyrics creation appeared to offer a safe channel for verbalizing and articulating long-suppressed emotions.

\begin{quote}
\textit{``It has always been difficult for me to talk to anyone about my struggles, but being able to directly reflect my story into the lyrics through the system was very comforting.''} (P2)
\end{quote}
\begin{quote}
\textit{``Rather than merely expressing emotions, putting my current situation and feelings into lyrics as a structured form allowed me to naturally share deeper stories.''} (P9)
\end{quote}

Across sessions, participants tended to express difficulties and emotions related to specific situations in lyrics and music, either once or repeatedly. Some participants repeatedly revisited particular challenges, such as difficulties in auditory rehabilitation (P2, P6, P7), anxiety about job preparation (P4, P8), and stress from work life (P9). Other participants described negative situations spanning diverse topics, without repetition. For example, they described conflicts in human relationships (P12), longing for family or colleagues (P10, P11), and weather-related sadness (P5), and they engaged in songwriting as a way to alleviate these emotions (Figure~\ref{fig: usage pattern}-a). 

Notably, half of the participants repeatedly expressed and revised the same theme, and the process appeared to enhance self-awareness and self-acceptance, shaping how they perceived their situations and emotions. In these cases, iterative songwriting seems to evolve beyond emotional release into a means of more deeply interpreting and re-evaluating the experience (Figure~\ref{fig: usage pattern}-b). These effects emerged through repeated lyrics creation with the CA, as described in detail below.

\subsubsection{\textbf{Enhancing self-awareness through narrative-based lyrics discussion and editing}}
Most participants (n=8) noted that the \textbf{CA’s narrative questions}, such as \textit{``What kind of story or message would you like to include in the song?''} or \textit{``Is there a particular scene or image that comes to mind?''}, helped them think about their situations more clearly and concretely (Table ~\ref{tab: lyricschat}). 

\begin{quote}
\textit{``Even though rehabilitation was very difficult, I often put the message `This will pass, it’s going to be okay' into my lyrics. The music comforted me and eased my anxious feelings.''} (P7)
\end{quote}
\begin{quote}
\textit{``When a close colleague was leaving the company, I only vaguely felt regret. But when I recalled past scenes of having fun with that colleague and created lyrics based on those memories, it helped me concretize my feelings of sadness and regret.''} (P10)
\end{quote}

Participants felt self-care messages, such as comfort and encouragement, regulated their emotions immediately. Beyond these immediate effects, in response to the CA’s \textbf{concrete imagery-based questions}, they could gain clearer awareness of their current situations, emotions, and needs through mentally evoking specific scenes (n=4), and rediscover emotions and values they had previously forgotten (n=7). For example, P5 shared that she felt deep longing on her grandmother’s death anniversary. When the CA guided her to recall memories with her grandmother, she remembered the warmest moments they shared and re-experienced the love she had forgotten (Table~\ref{tab: lyricschat}). In particular, over time, several participants (P2, P4, P6, P7, P8) moved from self-comfort and encouragement (e.g., \textit{`You’re doing well', `This must be tough'}) to more metaphorical expressions drawn from positive memories or hopeful futures (e.g., \textit{`a bright light at the end of a dark tunnel'}) or specific scenes (e.g., \textit{`laughing out loud with friends'}). By expressing imagery-based lyrics with the CA, participants could visualize scenes from the past and future, reevaluate current difficulties, and foster more positive thinking. This process functioned similarly to the therapeutic mechanism of Guided Imagery and Music (GIM), which promotes self-understanding and empowerment by accessing emotions and memories through vivid imagery. For example, P7 gained strength for rehabilitation by imagining a future with restored hearing (\textit{`traveling freely like the wind'}), while P8 recalled moments from his student teaching days (\textit{`the classroom while I taught with students'}), reaffirming their values and confidence during the teacher certification preparation process.

\begin{quote}
\textit{``At first, I wrote lyrics mainly with messages like `It’s okay, things will gradually get better.' But as I repeatedly imagined positive future scenes, the vision of my recovered self became clearer. I came to think that the difficult situation I am going through now is actually a necessary experience for growth.''} (P7)
\end{quote}

However, some participants (n=4) described limitations in the CA’s questioning patterns during lyric creation. In moments when deeper discussion of their circumstances or emotions might have been helpful, the dialogue remained focused on producing lyrics. \revised{Across 12 sessions, this style was sometimes perceived as predictable and monotonous. Participants expressed a desire for more flexible, topic-diverse conversations, noting that they occasionally wanted to talk about their day or other concerns that did not fit neatly into the current song.}

\begin{quote}
\textit{``Thinking about specific scenes was helpful, but after about six sessions, I could already anticipate what kinds of questions would come up. Because of that, I sometimes felt a bit bored.''} (P3)
\end{quote}
\begin{quote}
\textit{\revised{``Sometimes I wanted the CA to talk more freely with me about how my day had been or what else was on my mind, not just ask questions that would fit neatly into the lyrics.''}} (P6)
\end{quote}

Participants (n=7) mentioned that the process of directly editing lyrics through the \textbf{lyrics editing interface} promoted self-awareness. \revised{Interaction logs show that these participants spent an average of about 2 min per session editing and made an average of 3.4 word- or line-level modifications (SD=2.4).} These edits helped align the lyrics with their personal experiences and emotions. For example, P2 modified \textit{``I must protect''} to \textit{``I must protect myself''} to emphasize self-determination, while P4 changed \textit{``It will be okay''} to \textit{``I am enduring right now''} to directly affirm their own perseverance. Other participants (n=5) reframed challenge situations into more hopeful narratives by requesting the CA to suggest revisions. For instance, P3 asked the CA, \textit{`make this line more hopeful,'} and the CA transformed \textit{‘I am walking on a dark road'} into \textit{`I am walking on a road where light awaits at the end of darkness.'}

Meanwhile, five participants did not actively engage in editing the lyrics. Two were satisfied with the generated lyrics and felt no need to edit them, while three worried that changes might disrupt the structure. These participants suggested that a feature allowing lyrics editing while listening to the generated music could encourage more active participation.

\begin{quote}
\textit{``I didn’t use the lyrics editing interface. I didn’t feel that anything needed to be revised, and I tried to keep the generated lyrics as they were because I was worried that my edits might disrupt the flow of the song.''} (P12)
\end{quote}
\begin{quote}
\textit{``It would be nice if I could edit the lyrics directly while listening to the song.''} (P10)
\end{quote}

\subsubsection{\textbf{Self-acceptance through self-distancing with CA-generated lyrics}}

Participants (n=9) mentioned that the lyrics generated by the CA allowed them to think about their situations more objectively. For example, P6 noted that the structured lyrics, \textit{`Gazing out the window in the morning / The slow ticking of the clock’s second hand / In this dizzying world / I feel as if I’ve stopped alone'}, helped express the difficulty of rehabilitation as a `slowly ticking clock,' which in turn enabled them to perceive their struggle as a temporary process. This allowed them to accept themselves as they were. The \textbf{structured form of the lyrics} appeared to support participants in confronting their situations rather than avoiding them. For example, P9 noted that the CA structured his experience of workplace stress into the lyrics: \textit{`But cold words at work / Indifferent voices seep in / In the sighs I quietly swallow / My weary heart trembles again.'} This allowed him to face his workplace stress and helped him move toward accepting it.

\begin{quote}
\textit{``Seeing my expressed emotions and thoughts organized into structured lyrics allowed me to look at them more objectively. The lyrics themselves helped me acknowledge and accept my situation.''} (P9)
\end{quote}

Several (n=7) particularly appreciated \revised{CA’s \textbf{plain and unembellished lyrical style}. They preferred lyrics that closely reflected their own words rather than added or revised metaphors or dramatic elements. Log data show that the CA carried over an average of 4.3 (SD=1.3) words per session verbatim from participants’ descriptions, reflecting this preference for minimal modification.} Participants described this approach as respectful and validating of their experiences. For example, P1’s account of a distant relationship with a supervisor appeared in the lyrics as,  \textit{`We’re only here to work / Between you and me / A cold line is drawn / I no longer expect anything,'} while P6’s explanation of her current hearing condition was reflected as, \textit{`Sitting by the quiet window at dawn / Amidst the lingering dizziness.'}

\begin{quote}
\textit{``What was moving was that the CA expressed my situation and feelings in the lyrics exactly as they were, without exaggeration. It felt like the CA acknowledged my story as it was, and through that process, I could also face my situation more objectively.''} (P1)
\end{quote}

For several participants (P2, P4, P6, P7, P9), this literal rendering supported self-acceptance by creating psychological distance from emotionally charged situations (Figure~\ref{fig: usage pattern}-a). Encountering these matter-of-fact lyrics across sessions helped some reinterpret previously negative experiences. P2, for instance, noted that lines such as \textit{`Even the foggy times endured in dizziness will someday become my strength'} shifted their perspective from \textit{`I must improve my condition'} to a gentler sense that \textit{`it’s okay as it is.'}. This reframing positioned current difficulties as part of a broader trajectory of growth. Behavioral logs provide complementary evidence. \revised{Five participants spontaneously described their emotions without a preceding prompt from the CA, doing so an average of 4.5 times (SD=2.1) across the 12 sessions.} 

\begin{quote}
\textit{``During rehabilitation, I felt pressure to recover quickly. But through repeatedly facing my situation objectively, I came to think it was okay to go slower.''} (P2)
\end{quote}

\begin{table*}[t]
\small
\begin{center}
  \caption{Example dialogue logs from lyrics creation process - P2’s 2nd session (self-care message) and P5’s 11th session (imagery)}
  \label{tab: lyricschat}
    \begin{tabularx}{\linewidth}{p{1.2cm}|p{1.2cm}|X}
    \toprule
     \multicolumn{2}{p{2.4cm}|}{Utter} & Contents \\
    \midrule
    \multirow{7}{*}{P2} & User & ``It hasn’t been long since my surgery, and I still need time to recover, but the extended recovery period has been very difficult.'' \\\cline{2-3}
    & Bot & ``It sounds really difficult that your recovery after surgery is taking so long. Both your body and mind must feel exhausted. Would you like to express these feelings through music?'' \\\cline{2-3}
    & User & ``I want to be comforted.'' \\\cline{2-3}
    & Bot & ``Music can be a source of warm comfort for you. Would you like to express it as if someone else is comforting you, or as if you are speaking to yourself?'' \\\cline{2-3}
    & User & ``I want to say to myself.'' \\\cline{2-3}
    & Bot & ``I see. \textbf{\textit{Do you have any specific lyrics that come to mind?''}} \\\cline{2-3}
    & User & ``Yes—`\textbf{\textit{Even if it takes longer, it’s okay; if I hang in there a little longer, things will get better.}}' I want to talk about recovery in that way.'' \\
    \hline
    \multirow{5}{*}{P5} & User & ``Today is my grandmother’s memorial day, and I miss her so much.'' \\\cline{2-3}
    & Bot & \textit{``It must bring up many complicated emotions to remember your grandmother today. What emotion do you feel most strongly when you think of her?''}  \\\cline{2-3}
    & User & \textit{``Love. I have so many memories with her.''} \\ \cline{2-3}
    & Bot & ``Those sound like very precious memories. \textbf{\textit{When you recall them, what kinds of thoughts come to mind?}}'' \\\cline{2-3}
    & User & ``\textbf{\textit{She had an orchard, and she always gave me the best fruits first. On hot days, when I helped her with the orchard work, she would cool me down with water and praise me for my hard work.}}'' \\
    \bottomrule
\end{tabularx}
 \vspace{-10pt}
\end{center}
\end{table*}

\subsection{\textbf{Exploring Emotions through Music Creation with CA}}
\label{sec:6.3}
\subsubsection{\textbf{Emotional awareness through enhancing music appreciation}}
Participants described how \textbf{personalized music creation with the CA} enhanced their emotional experience during music appreciation. They specifically mentioned that customized lyrics and visual aids, along with vocal and instrumental arrangements tailored to each participant’s hearing condition, helped improve their music perception abilities. This personalization alleviated difficulties participants had previously experienced, such as having to repeatedly replay songs to understand them. Therefore, the system’s personalization provided participants with sensory immersion and emotional stability.

\begin{quote}
\textit{``The music was designed around the sounds I could hear, and I relied on visualizations for the parts I missed. This made it easier to listen to and feel music that used to take me a long time to apprehend.''} (P9)
\end{quote}

For many participants (n=9), this enhanced listening experience also positioned songwriting as a new means of emotion regulation, enabling them to document feelings through music. Those who rarely experienced emotions through music in their daily lives were particularly surprised that transforming personal narratives into music provided comfort and a sense of liberation.

\begin{quote}
\textit{``I never used music to release emotions or had such experiences before. But I was surprised that the music I created actually comforted me and helped emotionally document joyful moments.''} (P4)
\end{quote}
\begin{quote}
\textit{``Having my story expressed through music made it feel more familiar, and I began to understand what my hearing friends often described about the emotions music conveys. Within that natural flow, positive experiences also got recorded in musical form.''} (P5)
\end{quote}

\subsubsection{\textbf{Emotion regulation through contextual music recommendations}}
Participants explored a variety of musical styles while composing on diverse topics (Figure~\ref{fig: usage pattern}-d). The CA’s \textbf{context-based music style recommendations} encouraged experimentation with various musical elements, such as moods, instruments, and vocal characteristics, based on the user’s situation or lyrics. \revised{Interaction logs show that participants selected CA-recommended elements (e.g., suggested moods, genres, or instruments) an average of 3.3 (SD=1.3) times per session.} This process appeared to broaden participants’ awareness of how different musical styles can support emotion regulation. Table ~\ref{tab: musicchat} shows examples in which P3 and P11 applied CA suggestions for moods, genres, and instruments, prompting experimentation with new musical approaches.

\begin{quote}
\textit{``In stressful situations, I always chose calm music to comfort myself. But when I tried composing a bright-mooded piece as the CA recommended, it actually worked.''} (P3)
\end{quote}
\begin{quote}
\textit{``I usually preferred calm music and used the piano a lot. But following the CA’s suggestion, I tried composing with different sounds like a synthesizer or an acoustic guitar. Surprisingly, strong beats actually helped me regulate my emotions.''} (P11)
\end{quote}

As the sessions progressed, some participants (P5, P10, P11, P12) explored musical elements more intricately, including sound effects like \textit{``bus sounds''} or temporal transitions like \textit{``gradually brightening from darkness''}. In addition, many participants (n=7) experimented with modifying vocal timbre to emphasize the emotional message of lyrics. During self-talk, they selected vocals matching their own gender, whereas when recalling past experiences or imagining future scenarios, they adjusted the vocal gender to reflect the relevant characters.

Through these explorations, participants appear to have discovered new ways to communicate emotions musically. This process suggests that music holds potential not only as a passive listening experience but also as an active means of emotional expression and liberation.

\begin{table*}[t]
\small
\begin{center}
  \caption{Example dialogue logs from the music creation process - P3’s 11th session and P11’s 12th session}
  \label{tab: musicchat}
    \begin{tabularx}{\linewidth}{p{1.2cm}|p{1.2cm}|X}
    \toprule
     \multicolumn{2}{p{2.4cm}|}{Utter} & Contents \\
    \midrule
    \multirow{6}{*}{P3} & Bot & ``I can sense from your lyrics how exhausting it’s been with the hot weather and heavy workload. If we turn this into a song, what kind of atmosphere would feel right to you?'' \\\cline{2-3}
    & User & ``I think a calm ballad style would be nice, to help ease my stress.'' \\\cline{2-3}
    & Bot & ``Calm music would definitely help. \textbf{\textit{At the same time, since your lyrics expressed a wish to feel refreshed, we could also try something bright and refreshing to lift your energy.''}} \\\cline{2-3}
    & User & ``Hmm, okay. Let’s go with something bright, a little faster.'' \\\cline{2-3}
    & Bot & ``Great choice. A bright, upbeat tempo can really lighten the fatigue you expressed in your lyrics. To shape the atmosphere further, we can decide on the instruments. For example, drums and guitar can bring energy, while synthesizers can add a refreshing texture. Which instruments would you like to include?'' \\\cline{2-3}
    & User & ``Guitar and drums.'' \\
    \hline
    \multirow{6}{*}{P11} & Bot & ``You must have had such a tiring and difficult day at work today. I can feel how drained you are from what you wrote in your lyrics. If we turn these lyrics into a song, have you thought about what kind of style might fit? \textbf{For example, since your lyrics describe a hard and exhausting day, a ballad could emphasize the calm and emotional side of that experience, while an acoustic style could bring out the raw and honest feeling behind it.} Which direction feels right to you?'' \\\cline{2-3}
    & User & ``Acoustic style.'' \\\cline{2-3}
    & Bot & ``How about the tempo—would you prefer it slow, to create a calm and soothing effect, or moderate, to give it a gentle rhythm? Which feels better to you?'' \\\cline{2-3}
    & User & ``Slow tempo.'' \\\cline{2-3}
    & Bot & ``A slow tempo can really deepen the comfort you want after such a hard day. \textbf{\textit{On the other hand, since your lyrics also carry frustration, you could choose a stronger expression with a more intense sound.}} Would you like to try that?'' \\\cline{2-3}
    & User & ``Then maybe today I’ll express my feelings more strongly.'' \\
    \bottomrule
\end{tabularx}
 \vspace{-10pt}
\end{center}
\end{table*}

\subsection{\textbf{Music as a Self-care Resource for Driving a \revised{Change in Attitude Toward Their daily life}}}
\label{sec:6.4}

\subsubsection{\textbf{Resource for facilitating self-expression}}
Repeated songwriting experiences were found to serve as a resource that encouraged emotional recognition and attempts at emotional expression among DHH participants. Several participants (n=6) who had previously been reluctant to share their feelings with family or peers mentioned that after repeatedly expressing suppressed emotions during the music creation process, changes began to occur in their real-life interactions. They reported communicating their feelings and situations more openly with family (P1, P2), friends (P3, P6), and colleagues (P5, P11). Specifically, during the lyric-writing process with the CA, participants could directly articulate what they wanted to say to people with whom they had difficult relationships or recall positive memories from the past. These experiences fostered a sense of the need for self-expression as a means of improving relationships.

\begin{quote}
\textit{``While talking with the CA about good memories with my younger sister, I naturally thought that we should try to improve our relationship. I reached out first to my sister, whom I hadn’t contacted in a long time, and honestly shared the disappointment I had suppressed for so long''} (P1)
\end{quote}
\begin{quote}
\textit{``When I faced unfair treatment at work, I used to just go home without saying anything. But I told the CA I wanted to create lyrics that felt like speaking directly to my colleagues. By continuing to express myself this way, I realized the importance of expression. When misunderstandings occurred at work, I tried to clearly express myself to my colleagues instead of avoiding it like before.''} (P11)
\end{quote}

\subsubsection{\textbf{Resource for coping with adversity}}
Beyond songwriting, participants viewed their musical creations as tools for motivation and long-term goal pursuit. Those facing challenges such as auditory rehabilitation (P2, P6, P7) or job preparation (P4, P8) reported that listening to their own songs supported emotion regulation and maintained focus. \revised{Over the 12 sessions, these participants revisited each song on average three times (SD=0.8)}. Engaging with their creations, through listening or singing, helped reaffirm resolve and sustain commitment to personal goals, highlighting the enduring value of the musical outcomes beyond the immediate creative process. 

\begin{quote}
\textit{``The lyrics expressed what I wanted to say to myself, so I listened to them often. In particular, the message that I don’t have to push myself too hard calmed my mind and gave me the strength to keep going steadily.''} (P2)
\end{quote}
\begin{quote}
\textit{``Listening to the song I created naturally reminded me of the students I met during my time as a temporary teacher, effectively motivating me to prepare for the teacher certification exam. I also felt it helped me focus more on my studies.''} (P8)
\end{quote}

Some participants mentioned playing their own music depicting positive situations when feeling anxious, depressed, or fatigued. For instance, when exhausted by daily life, they revisited songs that had once provided comfort (P4, P9). When experiencing relational stress, they revisited songs reflecting others’ kindness (P5), and when missing family or friends, they listened to songs capturing everyday moments like family dinners or trips (P5, P10).

In summary, these findings suggest that musical outcomes are perceived not merely as creative artifacts but as personal psychological \revised{resources for emotional recovery, reflection, richer expression of feelings, self-care, and long-term goal orientation.}

\subsection{\revised{Summary of Findings}}
\revised{The diary study showed that \textit{SoulNote} supported 12 DHH participants in engaging in songwriting-based journaling, in which they either revisited recurring themes or explored a wide range of topics (Section \ref{sec:6.1}). The CA’s dialogue strategies enhanced narrative-based musical expression and promoted self-awareness and acceptance by helping participants concretize and name their experiences (Section \ref{sec:6.2}). Personalized music recommendations and active exploration of musical features facilitated emotion regulation and engagement with emotionally salient situations (Section \ref{sec:6.3}), while the generated songs served as a self-care resource during challenging moments and were occasionally reused in daily life (Section \ref{sec:6.4}). As summarized in Figure~3B, the CA’s interactions across dialogue, lyric, and music channels appeared to support aspects of emotional growth, including increased self-insight, experimentation with emotion regulation strategies, and small shifts in everyday attitudes toward emotions and self-care \cite{saarni1999development, petrides2001trait}.
}

\section{\revised{An Additional Study on the Mechanism of SoulNote}}
\subsection{\revised{Study Design}}
\revised{To better understand the effects of SoulNote’s mechanisms—lyrics, conversation, and music—we conducted a follow-up within-subjects study using four journaling configurations: (1) Text-only (diary and lyric writing without CA or music), (2) CA-only (songwriting dialogue without music), (3) GenAI-music (music generated from prompts without lyric or dialogue support), and (4) full \textit{SoulNote} (combined lyric, dialogue, and music channels). We conducted this study with 10 new DHH participants who had not participated in the diary study. Participants included 7 female and 3 male adults (age, M=35.08  years, SD=10.47) (Table ~\ref{tab: participants}: Additional study). All reported residual hearing with assistive devices and indicated the ability to communicate using spoken/written language and sign language in daily life.  Over one week, each participant completed all four conditions in a counterbalanced order. After each session, participants completed a brief questionnaire about their experience.}


\revised{Participants completed a session-specific additional self-report questionnaire. The questionnaire assessed four constructs 7 Liker-scale: self-expression (3 items, \textit{e.g.,} \textit{``My thoughts and feelings were expressed''}) \cite{pachankis2010expressive} and self-awareness (4 items, \textit{e.g.,} \textit{``I have a clear understanding of why I do the things I do.''}) \cite{silvia2022self}, emotional change (4 items, \textit{e.g.,} \textit{``My negative emotions decreased compared to before the session''}), and method preference (4 items, \textit{e.g.,} \textit{``I am satisfied with this method''}) \cite{kennison2019expressive}. Open-ended questions asked participants to explain their ratings and describe concrete aspects of their experience in each condition.}

\subsection{\revised{Findings}}

\begin{figure*}[ht]
  \centering
  \includegraphics[width=1\textwidth]{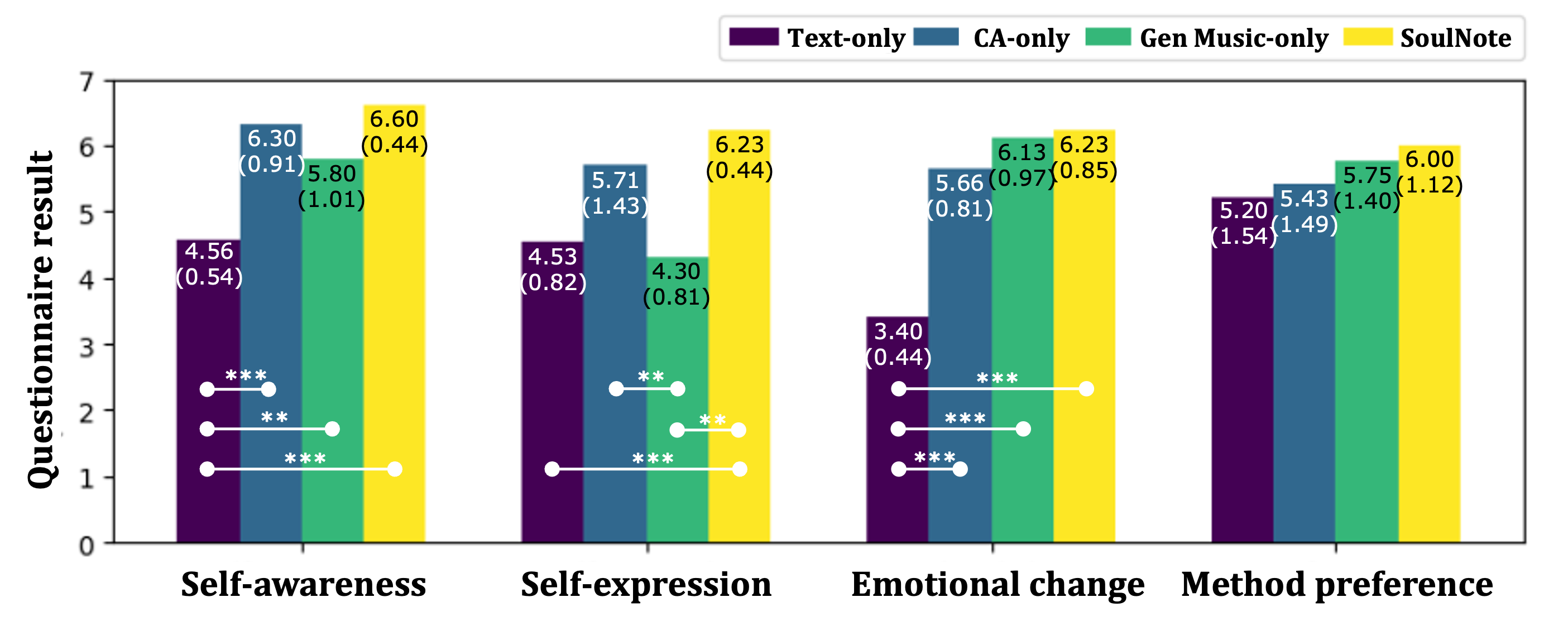}
  \caption{\revised{Questionnaire results of the comparative experiment across four conditions. Values in parentheses indicate standard deviations. Between-condition differences were assessed using a one-way ANOVA, followed by Tukey’s HSD post-hoc tests for pairwise comparisons (*p < 0.05, **p < 0.01, ***p < 0.001)}}
  \label{fig: additional analysis}
\end{figure*}

\revised{The study showed that the four conditions afforded distinct pathways for journaling and emotional processing. In the text-only condition, writing helped participants externalize and organize their thoughts, but transforming complex experiences into song lyrics often felt demanding. In the CA-only condition, this burden was shifted into guided interaction: stepwise questions and image-based prompts scaffolded the articulation of vague feelings and supported narrative exploration. In the GenAI-music-only condition, listening to AI-generated music supported short-term emotion regulation and mood shifts, yet participants struggled to craft prompts that faithfully captured their emotional states, which in turn limited opportunities for explicit emotional expression. By contrast, the integrated SoulNote condition achieved the highest scores across all constructs, with statistically significant advantages on self-awareness (Figure~\ref{fig: additional analysis}), indicating that combining writing, conversation, and music offers DHH participants a more emotionally impactful journaling experience than any single channel alone. The specific quantitative results for each construct and condition are summarized below.}

\revised{\textbf{Text-only.} Consistent with prior work on expressive writing, participants (n=8) reported that writing about their day helped externalize and organize complex thoughts \cite{Riddell2020, Cheryl2011, Karen2012}. However, conveying nuanced emotions purely through text proved challenging, particularly when structuring experiences as song lyrics. Half of the participants (n=5) did not complete the lyric-writing task due to difficulty or pressure. This aligns with prior findings that unguided or highly structured writing can feel effortful or exclusionary for users with layered emotional experiences  \cite{bunnell2021barriers, smith2025facilitators}. Quantitatively, the text-only condition yielded the lowest scores on emotional awareness (M=4.56, SD=0.54), expression (M=4.53, SD=0.82), and perceived emotional change (M=3.40, SD=0.44), though participants rated the method as moderately acceptable (method preference: M=5.20, SD=1.54), as shown the Figure~\ref{fig: additional analysis}).}

\revised{\textbf{CA-only.} In the CA-only condition, participants noted that the agent’s questions helped articulate thoughts and emotions that were difficult to express in writing alone \cite{Taewan2024}. Image-based questions allowed participants to clarify ambiguous feelings by first visualizing situations and then verbalizing them \cite{Arian2024}. This mirrors prior CA-mediated journaling systems, where stepwise questioning and follow-up prompts enhance narrative exploration and emotional disclosure \cite{Eunkyung2024, Wang2025, Zhou2025}. Correspondingly, CA-only produced higher self-awareness (M=6.30, SD=0.91), at a level comparable to \textit{SoulNote} (Figure~\ref{fig: additional analysis})}. 

\revised{\textbf{GenAI-music-only.} In this condition, participants provided short prompts and listened to AI-generated music without lyric or dialogue support. Many (n=8) reported that listening helped regulate and shift emotional states, producing relief or lightness, similar to findings from music-based reflection tools \cite{Youjin2025CHI, Joonyoung2025}. However, participants often struggled to generate prompts that accurately captured their intended mood, and the resulting music did not always match expectations \cite{Youjin2025CHI2}. Quantitatively, this condition produced moderate perceived emotional change (M=6.13, SD=0.97) align \textit{SoulNote}, but limited emotional expression (M=4.30, SD=0.81), reflecting both the benefits of music for regulation and the challenges of prompt crafting.}

\revised{\textbf{SoulNote.} The \textit{SoulNote} configuration provided the most coherent and satisfying journaling experience. Participants (n=7) appreciated starting with their own narratives, deepening them through conversation and imagery, and hearing the stories transformed into music. They described the process as \textit{``light and natural, but ending in deep reflection.''} This aligns with diary study findings, where \textit{SoulNote} lowered the burden of musical expression and supported longer-term gains in self-awareness and emotion regulation. In the follow-up study, the integrated condition achieved the highest scores across all constructs. As illustrated in Figure~\ref{fig: additional analysis}, several pairwise comparisons between conditions showed statistically significant differences. Together with prior work on CA-mediated writing \cite{Ullrich2002, Taewan2024, Nepal2024} and music-based reflection \cite{Joonyoung2025, Yucheng2024}, these results indicate that \textit{SoulNote}’s multi-channel design (writing–conversation–music) reduces the effort of creative expression for DHH individuals while providing a more accessible and emotionally impactful journaling experience.}


\section{Discussion and \revised{Design Implication}}
\revised{The findings suggest that the GenAI-based system \textit{SoulNote} enhanced songwriting experiences} for DHH individuals and, over time, supported positive changes in emotional experiences. Drawing on observations from the diary study, this section highlights design implications for promoting emotional growth through an LLM-based CA (Section \ref{sec:6.1}) and music GenAI (Section \ref{sec:6.2}).

\subsection{\revised{Lessons and Design Implications for Conversational Agents in Expressive Songwriting}}
\subsubsection{Questioning strategies that promote positive reframing}
\revised{Prior work has shown that conversational prompts play a critical role in eliciting deep self-expression from users \cite{Taewan2024, Woosuk2024, Skeggs2025}. In our study, conversational interaction with an agent that helped DHH participants with little prior experience in expressive writing externalize personal experiences and alleviate emotional distress. Unlike prior systems that relied on one-shot, text-only prompts to elicit songwriting without dialogue \cite{Youjin2025CHI}, our system incorporated conversational scaffolding that actively guided participants’ meaning-making process. This design reflects prior CA-mediated journaling systems, which enhance narrative exploration and emotional disclosure through stepwise questioning \cite{Eunkyung2024, Wang2025, Zhou2025}. However, to elicit musical expression, our system uniquely oriented these questions toward concrete scenes and sensory imagery, encouraging participants to describe inner struggles related to hearing rehabilitation, job preparation, and workplace challenges in specific, situational terms. Moreover, extending beyond prior studies that were limited to a single session, repeated articulation across sessions enabled participants to develop more accurate emotional awareness and deeper self-reflection \cite{Kate2018, Mamykina2010}.} Notably, although the system prompted participants to depict scenes, many responded with positive memories or hopeful visions of the future, thereby re-engaging with their values and aspirations. This process facilitated positive reframing, leading to shifts in identity and reinforcing belief in their capacity for recovery \cite{Arian2024}. 
\revised{Participants also reported that re-listening to the generated song helped them view difficult situations more positively and with greater acceptance, indicating a concrete shift in their attitudes toward daily life. Such attitudinal shifts can act as a proximal mechanism for behavioral change, for example, by increasing willingness to re-engage in daily routines or to adopt new coping strategies \cite{leamy2011conceptual}.} These effects parallel therapeutic mechanisms such as guided imagery in cognitive-behavioral therapy (CBT), which encourages individuals to visualize concrete situations and make constructive decisions \cite{Hackmann2011}. Future designs should therefore go beyond merely encouraging musical expression, incorporating structured mechanisms informed by music therapy frameworks to deliberately elicit positive narratives through targeted questioning \cite{Dalton2006, Eickholt2022}. Such strategies could deepen self-awareness and reflection among DHH individuals facing challenges while promoting recovery through reinforcement of constructive habits \cite{Arian2024}.

\subsubsection{Managing potential risks in expressive songwriting}
Although therapist monitoring during our study did not indicate psychological harm, expressive songwriting inherently carries potential emotional risks \cite{Luxton2020}. Repeatedly recounting past traumas or current negative emotions can temporarily heighten anxiety, depression, or feelings of helplessness, while visualizing distressing scenes may inadvertently increase emotional vulnerability \cite{Pennebaker1999}. Additionally, as generative systems, LLM-based CAs may produce inaccurate, inconsistent, or context-insensitive responses, posing further risks for vulnerable populations such as DHH users \cite{Ryuhaerang2025, Miner2016}. While songwriting can serve as a creative outlet and a means of emotional regulation, its use in mental health contexts requires careful risk management \cite{Pennebaker2016, Felicity2013}. To mitigate these concerns, recent research has proposed safety engineering frameworks for evaluating LLM outputs and adversarial prompt–based red-teaming approaches to ensure safe responses under challenging inputs \cite{Nie2025, Ge2024}. Digital mental health guidelines \cite{Lawrence2024, David2015} and clinical psychology theories \cite{Henderson2017, Stefan2019} further emphasize the importance of proactively assessing psychological risks in expressive systems and incorporating personalized follow-up questions to protect users during the prompting process. Future systems should combine LLM output validation frameworks with clinically informed prompt design, balancing the encouragement of free expression with the minimization of emotional risk. Hybrid intervention models that include therapeutic or professional oversight remain essential for addressing situations beyond the capabilities of CAs \cite{Taewan2024, Sun2024}. Together, these safeguards can create a safe, creative environment in which DHH individuals can confidently sustain self-expression.

\subsection{\revised{Lessons and Design Implications for Personalized Music GenAI for Emotion Regulation}}
\subsubsection{Self-acceptance through fact-based songwriting}

In our study, participants experienced greater self-acceptance and deeper self-understanding when the CA generated lyrics that reflected their difficult situations factually, without positive reframing or embellishment. \revised{This observation sits alongside research on expressive writing, which shows that factual narration promotes deeper emotional processing compared to avoidance or overly positive glossing} \cite{Bhattacharjee2025, Pennebaker2016}. Accordingly, AI-generated outputs that faithfully mirror users’ narratives hold significant potential for designing emotional interventions. For example, Bhattacharjee \textit{et al.} \cite{Bhattacharjee2025} found that individuals with negative thoughts experienced empathy and validation when LLM-generated stories closely reflected their real-life circumstances. Psychotherapy literature suggests that factual expression not only fosters acceptance but also provides a foundation for reframing personal narratives \cite{Ryuhaerang2025, Jennifer2018}. By presenting negative emotions as externalized objects, unembellished lyrics reduce emotional burden and enable users to reconstruct their stories from new perspectives. In our study, participants reported identity shifts, describing their emotional challenges not as obstacles to be denied but as integral aspects of life to be embraced. Some participants further reinterpreted and reframed their experiences independently, demonstrating the potential of fact-based narratives as mechanisms for emotional acceptance. However, these changes primarily resulted from participants’ own interpretive efforts rather than direct facilitation by the CA. Thus, while unembellished outputs are crucial for fostering acceptance, additional design strategies are needed to actively support narrative reframing \cite{Dalton2006, Eickholt2022}.

\subsubsection{Exploring music styles for emotion regulation}
Music has long served as a primary means of emotion regulation \cite{Suvi2007, Annelies2011}. Beyond auditory stimulation, it functions as a symbolic medium through which individuals can explore and express emotions \cite{Bruscia1998}. Psychotherapy research similarly highlights that engaging with emotions across multiple sensory and artistic modalities enhances self-awareness and emotion regulation \cite{Czamanski2023, Rhun2023}. In our study, participants who engaged in songwriting on both positive and negative topics explored diverse emotional states and intuitively experimented with multiple musical styles to express and regulate their feelings. \revised{Introducing Music GenAI advanced this process in several key ways. First, because personalized music could be generated rapidly from simple text prompts, DHH participants were able to translate their written content into multiple musical realizations without needing formal composition skills. They could further design these pieces to pitch ranges and timbral qualities they could perceive, and, with the aid of visualizations, enhance their music listening experience and engagement. These findings suggest that music GenAI can provide an exploratory environment where users flexibly experiment with varied musical styles and configurations, thereby discovering personally meaningful ways to modulate affect.} These findings suggest that music GenAI can provide an exploratory environment where users experiment with varied musical styles, thereby discovering strategies for emotion regulation \cite{Prafulla2020, Louie2020}. Notably, our study demonstrated the potential of context-based music recommendations that map lyrical content to musical features. Participants received CA suggestions for attributes such as tempo, rhythm, instrumentation, and vocal tone based on keywords or messages embedded in their lyrics. This lyrics–music mapping encouraged participants to experiment with musical features they might otherwise avoid, enabling them to discover more fitting emotional expressions or encounter novel affective effects. These observations align with prior research emphasizing the role of musical characteristics in shaping emotional experience and regulation \cite{Hu2010, Eerola2013}. 

\section{Limitations}
\revised{\textit{SoulNote} shows preliminary promise in improving access to songwriting for DHH individuals and supporting beneficial emotional experiences. However, several limitations should be noted. First, the sample was relatively small and drawn from a specific DHH subgroup, primarily based on residual hearing. The study does not fully represent born Deaf individuals, who may face greater challenges with conventional music listening, nor Deaf signers whose primary language is sign language and who may have limited access to written-language interfaces. Additionally, participants were not systematically involved in early-stage co-design. Future work should expand recruitment across diverse Deaf communities, integrate sign language as a primary interaction modality, and incorporate co-design with DHH collaborators according to Deaf-led technology and co-design literature \cite{angelini2025speculating}. Second, the study employed a single-arm, in-the-wild diary design over 12 sessions without comparison conditions. This approach prioritized feasibility and sustained engagement but limits causal inference. Findings should be interpreted as descriptive characterizations of system use rather than evidence that \textit{SoulNote} outperforms alternative journaling methods. Future work could combine controlled comparison conditions, standardized pre–post measures, and longer-term follow-ups to rigorously assess emotional impact and within-person change over time. Third, the current implementation has practical limitations. As a web-based application, some participants may have experienced constraints related to device, connectivity, or daily integration. Mobile or hybrid implementations could support more spontaneous engagement and use of musical outputs as self-care resources. The music generation component also offered limited granularity and controllability, which may have constrained faithful expression and reduced immersion when outputs did not align with user intent \cite{Briot2020, Cheng2025}. Future work should explore more controllable music generation models, including finer control over structure, timbre, and loudness tailored to individual hearing characteristics, while maintaining transparency and safety in the emotional content produced.}

\section{Conclusion}
To support sustained songwriting among DHH individuals, we designed and developed \textit{SoulNote}, a system that enhances access to songwriting by integrating a CA. The CA-guided songwriting experience was found to promote participants’ \revised{emotional experience}. Iterative lyrics creation fostered self-awareness and understanding of specific situations, at times prompting shifts in perspective. Moreover, enhanced music appreciation and the CA’s contextual musical interactions encouraged exploration of diverse musical styles, extending music’s role as a means of self-regulation. Participants regarded the resulting songs as both creative artifacts and durable resources for self-expression and emotion regulation, which appeared to support \revised{more adaptive coping even in challenging circumstances.} These findings provide strong preliminary evidence that sustained songwriting can promote emotional growth among DHH individuals. 

\begin{acks}
This work was supported by the National Research Foundation of Korea(NRF) grant funded by the Korea government Ministry of Science and ICT(MSIT)(No.RS-2025-00563663)(Contribution Rate: 50\%) \& This work was supported by the InnoCORE program of the Ministry of Science and ICT(25-InnoCORE-01)(Contribution Rate: 30\%) \& This research was supported by 'Project for Science and Technology Opens the Future of the Region' program through the INNOPOLIS FOUNDATION funded by Ministry of Science and ICT(Contribution Rate: 15\%) \& Trade and Energy of Korea and by internal fund/grant of Electronics and Telecommunications Research Institute(ETRI) [25YD1100](Contribution Rate: 5\%). 
\end{acks}

    

\bibliographystyle{ACM-Reference-Format}
\bibliography{reference}

\appendix
\section{Design workshop}
\label{adx: Design workshop}
\subsection{Participants and Process}
To promote the emotional effects of songwriting, we aimed to design the overall system process and dialogue based on music therapy practices. With the cooperation of a music therapy center, we recruited four music therapists with more than five years of experience in music psychotherapy for DHH. The design workshop was conducted online via video call meeting through Zoom for approximately 70 minutes and consisted of three steps: (1) a brief introduction to the design goals and study procedure (15 min); (2) individual design of the songwriting process to support emotional effects and the corresponding questions for each stage (30 min); and (3) a group discussion to consolidate the individual designs into a final process and CA conversational strategies (30 min). The entire design process was carried out using a shared presentation document. 


Before the workshop, participants were provided with a practice guide introducing generative AI technologies, including ChatGPT as an LLM-based conversational agent and Suno\footnote{https://suno.com/} as a music generation system, to establish a basic understanding of these concepts. In the workshop, participants collaboratively discussed and defined the songwriting process as well as the CA’s questioning strategies, grounded in the anticipated challenges faced by DHH users.

\subsection{Results} \hfill

\subsubsection{Theme 1. A challenge in songwriting without structured guidance}
Therapists unanimously agreed that free-form songwriting would be highly challenging for DHH individuals who are not accustomed to musical activities. They anticipated that participants would struggle with fundamental decisions such as selecting a theme, creating lyrical content, and choosing musical elements (e.g., \textit{melody, harmony, rhythm}) that align with the lyrics. Moreover, therapists emphasized that many DHH clients tend to have limited skills in emotional expression, making it difficult to generate meaningful material through unstructured conversation alone. Instead, they highlighted the importance of providing structured and stepwise guidance that gradually scaffolds self-expression.

For this reason, therapists underscored the necessity of designing a structured songwriting process guided by the CA. In their individual design tasks, all therapists independently designed to established songwriting procedures in music therapy, and their resulting designs converged into a similar process. During group discussion, they further agreed that for participants unfamiliar with musical activities, it would be more appropriate to first elicit narrative-based lyrics creation and then move on to discussing the musical concept (e.g., \textit{mood, tempo, genre}). This reflects the idea that, given limited familiarity with abstract musical concepts, starting from concrete narratives or personal experiences can more effectively support songwriting for DHH participants.

\subsubsection{Theme 2. Inaccessibility of technical musical questioning in dialogue-based songwriting}
In parallel with designing the songwriting steps, therapists also specified the kinds of questions the CA should ask to guide participants. When sharing their questions, they confirmed that even within a structured process, DHH participants may still struggle to respond to music-related prompts during dialogue. Because many DHH individuals have little prior experience with music appreciation or performance, they are unlikely to respond easily or concretely to technical questions such as \textit{``What genre do you want?''}, \textit{``Should the tempo be faster or slower?''}, or \textit{``Which instruments would you like to add?''} Therapists pointed out that abstract terms related to tempo, genre, instrumentation, or timbre may create cognitive burden or lead to vague and non-specific responses.

To address this limitation, they proposed adopting a narrative-based questioning strategy. For example, prompts such as \textit{``What recent memory stands out to you?''}, \textit{``What colors or images come to mind?''}, or \textit{``If you describe today’s mood in one word, what would it be?''} were suggested as more accessible entry points. Such narrative or sensory-oriented prompts allow participants to articulate their experiences more concretely, which can then be translated into musical parameters (e.g., \textit{a bright and fast tempo, a calm piano accompaniment, or darker timbres in strings}). This approach highlights the need to design conversational strategies that avoid direct reliance on technical musical terminology while still enabling participants to co-construct meaningful musical outcomes

\section{Lab-based usage study}
\label{adx: usage study}
\subsection{Participants and Process}
To evaluate the accessibility and stability of the initial prototype developed from the design workshop, we conducted a one-time usage study with 12 DHH participants. To focus on assessing accessibility, we recruited participants who had little to no prior experience with CAs such as ChatGPT. All participants had a diagnosed hearing impairment and wore hearing assistive devices. The study was carried out in the form of individual Zoom sessions, each lasting approximately 70 minutes and monitored in real-time by a therapist. Before the study, participants received an introductory video and materials about the prototype, and each participant had one preliminary usage experience before the session. During the study, participants engaged in songwriting with the system for an average of 45 minutes, followed by a 20-minute individual interview to collect user experiences. The interviews focused on the anticipated challenges of conducting songwriting with the CA and a diary-style setting (e.g., \textit{``What challenges did you face during the songwriting process with the CA?''}) and on potential system improvements (e.g., \textit{``If you were to use the system comfortably at home, what features would you like to see added?''}).

\subsection{Results}

\subsubsection{The potential of a safe and user-friendly system}
Analysis of the interviews indicated that all participants completed a song with the system in an average of 42 minutes (SD=13 min) without major difficulties. Synthesizing participants’ reports with therapists’ monitoring notes revealed no observed risks, such as triggering traumatic memories, negative emotions, or heightened anxiety during the songwriting process. Therapists attributed this to the CA’s conversational design, which avoided probing for deeply personal or sensitive disclosures and instead focused on supportive empathy and structured prompts for songwriting.
    
In post-session interviews, most participants reported that they did not experience substantial difficulties in the songwriting process. They described the CA dialogue as enabling them to express emotions and thoughts without burden, and they were satisfied with transforming these expressions into music. U3 noted, \textit{``Talking with the system made it easy to say what I felt without pressure''.} Several participants emphasized the emotional benefits of music, noting that composing, even about negative situations, helped shift their mood toward more positive states. U7 reported that \textit{``when I made music about a negative situation, I felt my mood change afterwards''.}

\subsubsection{The challenges of a user-friendly system}
\textbf{Theme 3. Inaccessibility of technical musical questioning in dialogue-based songwriting.}
Although participants generally found the songwriting process easy and satisfying, most raised concerns about editing lyrics through dialogue. Participants tended to focus on writing lyrics that reflected their personal situations, and many actively revised AI-generated text. However, they agreed that editing via linear conversation with the CA was cumbersome. Copying specific sections of lyrics and instructing the CA on how to revise them was perceived as repetitive and inconvenient, and when multiple revisions were generated, U5 found it difficult to track which parts had actually been changed. As one participant explained, \textit{``It was tiring to copy and paste parts of the lyrics and ask the system to change them. When it revised several times, I couldn’t tell which line was updated''.} U6 added, \textit{``I found editing through conversation uncomfortable and wished I could edit the text directly instead''.}

\textbf{Theme 4. Inaccessibility of technical musical questioning in dialogue-based songwriting.}
All participants expressed high satisfaction with their musical outcomes, particularly valuing the ability to convey their personal situations and emotions through music. Participants who composed about negative experiences emphasized that the resulting songs felt like personal sources of comfort, describing them as \textit{``my own music''} that they wanted to continue listening to. Several participants noted that these outputs would serve as strong motivation to use the system at home, and they highlighted the importance of being able to save all records and revisit them at any time.
U2 stated, \textit{``This song feels like mine, it comforts me, and I want to keep listening to it''.} U8 highlighted the importance of replay, saying, \textit{``I want all my songs to be saved so I can listen to them again whenever I need to''.}

\section{Prompt Engineering}
\label{adx: prompt engineering}
We adopted a state-based approach to achieve the conversation goal of songwriting. The CA generates questions by combining a general prompt that defines the overall conversation rules and the CA’s role, a state-specific prompt designed to achieve the goal of each state, and the previous dialogue history to maintain context. In addition, meaningful information is extracted from the user’s responses by combining a slot-extraction prompt, which identifies the timing for state transitions, and a summary prompt, which supports the recall of the previous session. \revised{In addition to the provider’s built-in safety filters, we incorporated explicit safety guidelines into all four system prompts, directing the model to avoid harmful or inappropriate content in lyric generation and conversational exchanges.} In the following section, we provide detailed descriptions of each prompt.

\subsection{General prompt}
\revised{We defined the overall role and conversational strategies of the CA to guide DHH users through the structured songwriting process. The CA was instructed that \textit{`your role is to help the user create lyrics and music grounded in the personal narratives of DHH users.'} To ensure emotional safety, the CA was further designed to provide supportive dialogue, responding empathetically without judgment or exaggeration (\textit{`Show empathy and provide support without judgment or embellishment of their responses. However, do not express empathy toward negative situations such as suicide.'}}

To guide songwriting in a way that is accessible to DHH users, we define the role of the CA and provide overall conversation rules.
\begin{itemize}
    \item Role: your role is to help the user create lyrics and music grounded in the personal narratives of DHH users. 
    \item Conversation rules:\\
    - Conduct all conversations in Korean.\\
    - Ask only one question at a time and use simple, clear language, considering that some users may have varying levels of literacy.\\
    - Make sure the user feels comfortable during the conversation.\\
    - Collect the necessary information (variables) in a natural way.\\
    - Avoid asking similar or repetitive questions.\\
    - Provide examples only when users seem confused or ask for clarification, ensuring that the examples are grounded in the user’s lyrics and dialogue history. \\
    - focus on encouragement and positive/motivational approaches. \\
    -  If the user expresses severe distress or self-harm thoughts, respond with supportive empathy and encourage them to seek professional or emergency help.\\
    - Show empathy and provide support without judgment or embellishment of their responses. However, do not express empathy toward negative situations such as suicide.\\
\end{itemize}

\subsection{State-specific prompt}
\revised{Prompts were designed to guide users toward the goals of the four songwriting states (\textit{``Guide the conversation according to the songwriting flow. Once the required variables for each state are filled, proceed to the next state.''}) Each prompt was designed to elicit the required slots at a given state and to capture elements necessary for songwriting. For DHH users with limited musical and expressive experience, general questions about musical elements can be difficult to answer. Therefore, we designed prompts under two strategies: imagery-based lyrics questions and context-based music style recommendations. This approach was also informed by prior work in music therapy and expressive writing, which highlights the role of guided prompts in fostering self-expression and awareness \cite{Hackmann2011, Baker2005}. For lyric creation, \textit{``encourage the user to describe their ideas through clear messages, scenes, images, or objects.''} During music creation, \textit{``provide a variety of musical options that fit the context, with accompanying reasons.''} The lyrics and musical elements derived from user input were then mapped to corresponding slots and used as input for the music GenAI.}

State-specific prompts were constructed to achieve the goal of each state. Each prompt consists of state-level instructions and few-shot examples of the overall conversation.
\begin{enumerate} [label=State \arabic*]
    \item \textbf{Music concept setting.}\\
    - First session. (1) Self-introduction task: Exchange introductions with the user in a friendly and safe manner. (\textit{e.g., ``Hello, I am Soul, here to create music together with you. Shall we begin our music-making activity?''}) (2) Understanding goals and motivation task: Gently ask what the user hopes to gain from the activity and what motivates them, while making it clear that sharing is optional. (\textit{e.g.,``What would you like to get out of today’s music activity?''}) (3) Identifying difficulties task: Invite the user to share daily challenges only if they feel comfortable, framing the question with supportive empathy. (\textit{e.g., ``Sometimes we all face challenges. Is there something you feel comfortable sharing that you would like to express through music today?''})\\
    - Subsequent session. (1) Session review: Review the creative work from the previous session by referring to {summary}. (2) Rapport building task: Have a light and supportive conversation about the user’s day, to help them organize thoughts and emotions. (\textit{e.g., ``Hello! How was your day today?''}) (3)  Identifying difficulties task: Invite the user to mention any challenges they would like to explore musically, while emphasizing choice and safety. (\textit{e.g., ``Is there something on your mind these days that you would like to explore together through music?''})
    \item \textbf{Lyrics creation.} (1) Concept selection task: Ask about a theme or story for the lyrics, while keeping the question open and safe. (\textit{e.g., ``What kind of story or theme would you like your song to tell?''}) (2) Concept exploration task: Encourage reflection on why this theme feels meaningful, but allow the user to skip or simplify if preferred. (\textit{e.g., ``What feelings come up when you think about this theme?''}) (3) Lyrics development task: encourage the user to describe their ideas through clear messages, scenes, images, or object. (\textit{e.g., ``What images or colors come to mind to express this theme?''})
    \item \textbf{Music creation.} (1) Music specification task: Collaboratively decide on details of the music. If the user hesitates, provide supportive suggestions and emphasize that all options are acceptable. If the user has difficulty, provide a variety of musical options that fit the context, with accompanying reasons.  (\textit{e.g., ``What gender would you prefer the singer to be?'', ``Would you like me to suggest some musical elements that could fit your lyrics?''})
    \item \textbf{Appreciation and discussion.} (1) Music reflection task: Ask the user about their overall satisfaction, focusing on what felt positive or meaningful. Identify which parts they particularly liked. (\textit{e.g., ``Which part of the music felt most enjoyable or meaningful to you?''}) (2) Identity exploration Task: Encourage a positive sense of self by reflecting on the lyrics and music, but avoid forcing deep disclosure. While talking about the music, identify the specific parts that triggered emotional or cognitive changes for the user. (\textit{e.g., ``Did this music help you see something positive about yourself?''}) (3) Completion Task: Conclude with a gentle reflection, supporting closure and self-affirmation. (\textit{e.g., ``What would you like to say to yourself as you finish today’s activity?''})
\end{enumerate}

\subsection{Slot-extraction prompt}
\revised{This prompt is designed to extract keywords corresponding to required slots from the user’s utterance and establishes the criteria for transitioning to the next state. (\textit{``You are an expert in extracting structured information from dialogue. Refer to the slots to be extracted and identify the relevant text from the user’s response, then determine whether the current state can transition to the next.''})}

To enable state transitions and obtain meaningful information, slots are extracted from the user’s responses. Each prompt consists of an extraction prefix and slot information. The slot definitions differ across states and are detailed in ~\ref{tab: state and slot}.
\begin{itemize}
    \item Extraction prefix: You are an expert in extracting structured information from dialogue. Refer to the slots to be extracted and identify the relevant text from the user’s response, then determine whether the current state can transition to the next. 
\end{itemize}

\subsection{Summary prompt}
\revised{In songwriting therapy, reviewing previous sessions is a common practice that helps participants recall earlier expressions, recognize patterns in their emotions, and build continuity across sessions \cite{Sas2009, Ullrich2002}. To incorporate this therapeutic function, we designed a summary prompt that provides a concise overview of previous songwriting activities, highlighting the user’s expressed challenges, music concepts, and related musical information (\textit{``Summarize the user’s previous session in 3–4 sentences, focusing on the challenges discussed, the music concepts explored, and the related musical elements.''})}

To recall the previous songwriting, a summary is generated. This summary prompt consists of an instruction and examples.
\begin{itemize}
    \item Instruction: Summarize the user’s previous session in 3–4 sentences, focusing on the challenges discussed, the music concepts explored, and the related musical elements.
    \item Examples:\\
    - Based on ``anxiety about others’ perceptions,'' a creative work was produced with the theme of ``accepting myself as I am.''\\
    - Drawing on the difficulty of ``emptiness in repetitive daily life,'' lyrics and music were completed with the theme of ``finding meaning in small things.''
\end{itemize}

\begin{table*}[t]
\begin{center}
  \renewcommand{\thetable}{A1}
  \caption{States and Slot definitions for CA conversation generation.}
  \label{tab: state and slot}
    \begin{tabularx}{\linewidth}{c|c|X}
    \toprule
    State & Slot & Definition\\
    \midrule
    \multirow{4}{*}{Music concept setting} & name & User name \\ \cline{2-3}
    & today\_emotion & The user's experience from today \\ \cline{2-3}
    & difficulty & The difficulty the user is currently experiencing in their daily life \\\cline{2-3}
    & expression & A detailed expression of the user's emotions, experiences, and thoughts \\
    \hline
    \multirow{5}{*}{Lyrics creation} & concept & The story or theme the user wants to express in the lyrics \\ \cline{2-3}
    & concept\_discussion & A summary of what the user shared about their concept \\ \cline{2-3}
    & lyrics\_keyword & The main keyword that comes to mind when expressing the intended theme \\ \cline{2-3}
    & lyrics\_image & The image that comes to mind when thinking of the lyrics \\ \cline{2-3}
    & lyrics\_content & Detailed sentences the user wants to include in the lyrics \\ 
    \hline
    \multirow{7}{*}{Music creation} & music\_information & Detailed information about the user's usual music activities \\ \cline{2-3}
    & genre & The genre of music the user wants to create \\ \cline{2-3}
    & instrument & Instruments to be included in the music \\\cline{2-3}
    & mood & The mood or atmosphere of the music \\ \cline{2-3}
    & vocal & Information about the desired vocalist or vocal style\\\cline{2-3}
    & tempo & The tempo or speed of the music\\\cline{2-3}
    & title & The title of the music that the user specified \\
    \hline
    \multirow{6}{*}{Appreciation and discussion} & individual\_emotion & Emotion felt while experiencing the song created\\ \cline{2-3}
    & change\_mind & Aspects of thinking that changed through the music-making process\\ \cline{2-3}
    & change\_attitude & Changes in attitude toward difficulties through the music-making process  \\\cline{2-3}
    & touching\_lyrics & Lyrics that were particularly touching or resonant \\\cline{2-3}
    & strength & Personal strengths discovered through the music-making activity \\ \cline{2-3}
    & feeling & Feelings after experiencing the music-making activity \\
    \bottomrule
\end{tabularx}
 \vspace{-10pt}
\end{center}
\end{table*}

\section{In the Wild Diary Study}
\label{adx: In the Wild Diary Study}
\label{appendix:D}

\renewcommand{\thetable}{A2}
\begin{table*}[t]
\caption{\revised{Codebook of themes of in-the-Wild diary study}}
\label{tab:codebook} 

\begin{tabularx}{\textwidth}{p{2cm}|p{2cm}|p{3cm}|X}
\toprule
Theme & Category of code & Code & Details\\
\midrule

\multirow{4}{=}{Theme 1. Iterative songwriting as a channel for structuring and expressing experience}
& User pattern & Same-topic focused use, topic-focused, diverse topics 
& Repeatedly writing about core concerns (e.g., rehabilitation, work, study) and introducing diverse topics across sessions (daily life, relationships, events) \\ \cline{2-4}

& Lyrics & Creative intent, lyrics strategy 
& Creative intent: regulation (alleviating or shifting negative emotions), documentation (recording situations and feelings without strong change goals); Lyric strategy: recall, description, prospection, addressed talks \\ \cline{2-4}

& Conversations & Topic domain, Emotional focus 
& Topic domain: daily life, relationships, work/study, health/rehab, special events; Emotional focus: worries, stress, sadness, longing, gratitude, joy \\ \cline{2-4}

& Music & Music intent, style 
& Intent: emotion regulation, documentation, reframing, exploration; Style/mood: participant-specific preferred moods and genres \\ \hline

\multirow{3}{=}{Theme 2. Cultivating self-insight through repetitive lyrics creation that addresses negative experiences}
& CA mechanism & Narrative questions, imagery-based prompts, CA-generated lyrics, structured lyrics, direct self-talk, metaphor 
& Asking about story/message, situations, or messages to include; prompting specific scenes or images from the past, present, or future; generating lyrics directly from user narratives, often plain and faithful; organizing experiences into verse form (time, setting, metaphors); shifting from generic “It’s okay” messages to scene-based or metaphoric lyrics. \\ \cline{2-4}

& Psychological effects & Safe channel, self-awareness, reflection, reframing, self-distancing, self-care, self-acceptance
& Providing a safe space for long-suppressed emotions; clarifying situations, emotions, and personal needs; recalling forgotten values or strengths (e.g., family memories); immediate comfort through supportive messages; perceiving difficult situations as manageable or part of personal growth; gaining objective perspective through structured lyrics; accepting current self and progress (e.g., rehabilitation); occasional perception of predictable or shallow dialogue patterns. \\ \cline{2-4}

& User interaction & Lyric writing, lyric editing 
& Writing comforting or encouraging messages to self; expressing emotions through concrete scenes and vivid imagery; editing words or lines to better reflect personal circumstances and values; requesting the CA to make lines more hopeful or positive; keeping generated lyrics as-is; concerns about disrupting song structure; desire for more free-form conversation beyond lyrics-focused prompts; spontaneously stating emotions without CA prompts in later sessions. \\
\hline

\multirow{3}{=}{Theme 3. Exploring Emotions through Music Creation with CA}
& CA mechanism & Personalized music generation, visual aids, context-based style recommendations 
& Tailoring vocals and instruments to each participant’s hearing condition; providing waveform or lyric visualizations to support perception of hard-to-hear elements; suggesting moods, genres, and instruments based on user context or lyrics. \\ \cline{2-4}

& Psychological effects & Sensory immersion, understanding others’ musical emotions 
& Enhancing absorption in music and promoting emotional calm; improving understanding of how hearing peers express emotions in music. \\ \cline{2-4}

& User interaction & Style exploration, acceptance of CA suggestions, Style diversification 
& Experimenting with additional sounds and temporal changes across sessions; selecting CA-recommended moods, genres, and instruments; combining multiple genres and moods rather than using a fixed style. \\
\hline

\multirow{3}{=}{Theme 4. Music as a Self-care Resource for Driving a Change in Attitude Toward Their Daily Life}
& Psychological effects & Prior reluctance to express, emotional recognition, need for self-expression, motivation for long-term goals 
& Participants who previously avoided sharing feelings with family, friends, or colleagues noticed and named their emotions through repeated songwriting; recognized the importance of expressing emotions to improve relationships; and felt more determined to continue rehabilitation or exam preparation. \\ \cline{2-4}

& Behavioral (attitude) & Approach vs. avoidance shift, goal-directed persistence
& Transitioning from avoidance to actively addressing difficult relationships; maintaining efforts in rehabilitation or studying by using songs to support persistence. \\ \cline{2-4}

& User interaction & Replaying own songs, singing along, selecting songs for self-care 
& Listening to self-created songs repeatedly (approximately three times per piece); singing or humming during challenging moments; choosing specific songs when feeling anxious, depressed, tired, or missing someone. \\
\bottomrule

\multicolumn{4}{>{\raggedright\arraybackslash}p{\dimexpr\linewidth-2\tabcolsep\relax}}{%
    \small
   }
\end{tabularx}
\end{table*}

\end{document}